\documentclass[11pt]{amsart}

\usepackage[utf8]{inputenc}
\usepackage[T1]{fontenc}
\usepackage{lmodern}

\usepackage{amssymb}
\usepackage{mathtools}
\usepackage{amscd}
\usepackage[mathscr]{eucal}
\usepackage{enumitem}
\usepackage[hmargin=1in,vmargin=1in]{geometry}

\usepackage{graphicx}
\usepackage{xcolor}
\usepackage[roman]{sublabel}

\usepackage[labelfont=bf,font=small]{caption}
\usepackage{subcaption}

\usepackage[noadjust]{cite}
\usepackage[foot]{amsaddr}

\usepackage{placeins}
\usepackage{bm}
\usepackage{siunitx}

\definecolor{cite_blue}{rgb}{0,0,1}
\definecolor{link_red}{rgb}{0.75,0,0}
\usepackage[colorlinks=true,
            citecolor=cite_blue,
            linkcolor=link_red,
            urlcolor=cite_blue]{hyperref}

\newlength{\arrayrulewidthOriginal}

\usepackage{ottmacros-shear-jmb}

\newcommand{\hmi}{{H^-}}
\newcommand{\hp}{{H^+}}
\newcommand{\hpm}{{H^{\pm}}}
\newcommand{\thp}{{\theta_+}}
\newcommand{\thm}{{\theta_-}}
\newcommand{\LLR}{{\rm LLR}}

\theoremstyle{plain}

\theoremstyle{definition}

\begin{document}

\raggedbottom
\thispagestyle{empty}

\title{Correlated Information Reduces Accuracy of Pioneering Decision-Makers}
% \title{Reliability failure in physiological systems}

\author{Megan Stickler$^{1}$}
\address{$^{1}$Department of Mathematics, University of Houston, Houston, Texas, USA}

\author{William Ott$^{*,1}$}
\thanks{$^{*}$\href{mailto:william.ott.math@gmail.com}{william.ott.math@gmail.com} (Corresponding author)}

\author{Zachary P. Kilpatrick$^{*,2}$}
\address{$^{2}$Department of Applied Mathematics, University of Colorado-Boulder, Boulder, Colorado, USA}
\thanks{$^{*}$\href{mailto:zpkilpat@colorado.edu}{zpkilpat@colorado.edu} (Corresponding author)}

\author{Kre\v{s}imir Josi\`{c}$^{*,1,3}$}
\address{$^{3}$Department of Biology and Biochemistry, University of Houston, Houston, Texas, USA}
\thanks{$^{*}$\href{mailto:kresimir.josic@gmail.com}{kresimir.josic@gmail.com} (Corresponding author)}

\author{Bhargav R. Karamched$^{*,4,5,6}$}
\address{$^{4}$Department of Mathematics, Florida State University, Tallahassee, Florida, USA}
\address{$^{5}$Institute of Molecular Biophysics, Florida State University, Tallahassee, Florida, USA}
\address{$^{6}$Program in Neuroscience, Florida State University, Tallahassee, Florida, USA}
\thanks{$^{*}$\href{mailto:bkaramched@fsu.edu}{bkaramched@fsu.edu} (Corresponding author)}

% \urladdr[William Ott]{http://www.math.uh.edu/$\sim$ott/}

%\keywords{}

%\date{\today}

\begin{abstract}
%Our decisions are shaped by observations of our environment, as well as the people around us. 
Normative models are often used to describe how humans and animals make decisions.
These models treat deliberation as the accumulation of  uncertain evidence that terminates with a commitment to a choice.
When extended to social groups, such models often assume that individuals make independent observations.
However, individuals typically gather evidence from common sources, and their observations are rarely independent.
Here we ask:
For a group of ideal observers who do not exchange information, what is the impact of correlated evidence on decision accuracy?
We show that even when agents are identical, correlated evidence causes decision accuracy to depend on temporal decision order.
Surprisingly, the first decider is less accurate than a lone observer.
Early deciders are less accurate than late deciders.
%Early decisions are more likely to  be based on faulty evidence when agents use the same decision criterion, which does not occur when  observations are independent.
These phenomena occur despite the fact that the rational observers use the same decision criterion, so they are equally confident in their decisions.
%Rational early and late deciders are equally confident in their decisions even when the accuracy of these decisions differs considerably.
We analyze discrete and continuum evidence-gathering models to explain why the first decider is less accurate than a lone observer when evidence is correlated.
Pooling the decisions of early deciders using a majority rule does not rescue accuracy in the sense that such pooling results in only modest accuracy gain.
%Pooling early decisions does not necessarily resolve this problem, since early agents' decisions can be correlated.
Although we analyze an idealized model, we believe that our analysis offers insights that do not depend on exactly how groups integrate evidence and form decisions.
%Although we describe an idealized, tractable setting, we argue that similar effects could impact group decisions generally.
\end{abstract}

\maketitle

\section{Introduction}

\begin{figure}[h!]
\includegraphics[width = \textwidth]{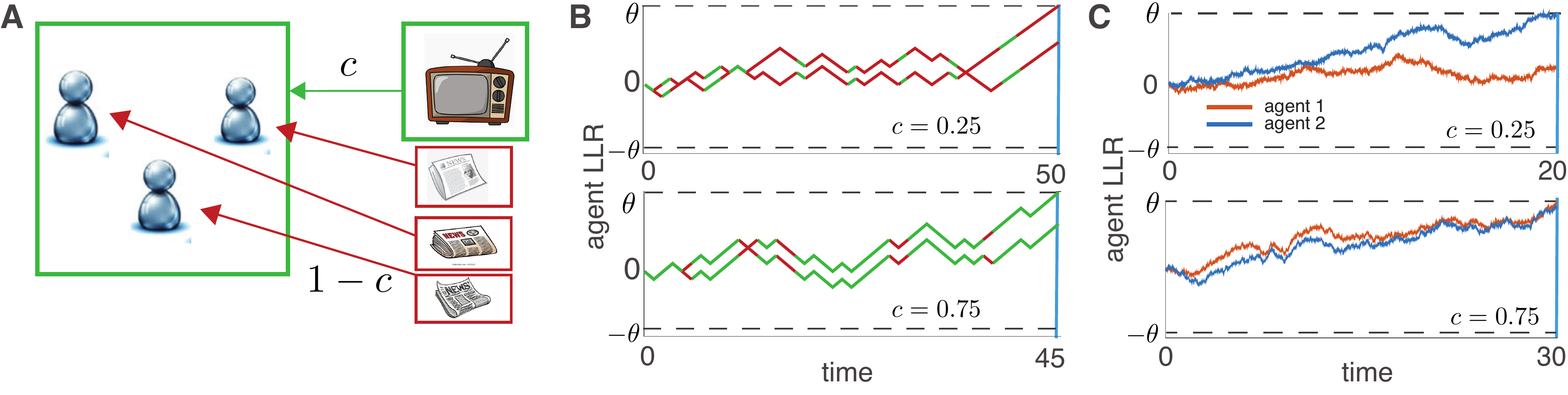}
\caption{Agents receiving partially correlated evidence. (a) Agents make a sequence of measurements to decide between alternatives. At each timestep, agents all make the same observation with probability $c$, and independent observations  with probability $1-c$. (b) Representative trajectories of a discretely computed log-likelihood ratio (LLR), Eq.~\eqref{E:corr_eq}, for $c = 0.25$ and $c = 0.75$ and two agents.  Green segments correspond to increments due to common observations and red segments arise from independent observations. An agent commits to a decision ($\pm$) when the computed LLR crosses $\pm \theta$.  (c) Analogous trajectories generated using the limiting drift-diffusion model, Eq.~\eqref{E:macroscopic}.}
\label{fig1}
\end{figure}

Most organisms and many computational algorithms make decisions based on a sequence of noisy observations of the environment~\cite{Gold02}. Normative models that describe how evidence should be integrated to make the best choice are central to our understanding of such decisions~\cite{Bogacz2006}. When an observer needs to choose between alternatives, accumulating evidence refines their perceived probability of the truth of each alternative. Decision policies often prescribe a threshold on the accumulated evidence in order to balance the speed and accuracy of decisions~\cite{chittka2009speed,bogacz2010}. These theories have been developed and validated over decades in experiments with humans and other animals~\cite{Ratcliff1978theory,chittka2003bees,newsome1989neuronal,uchida2003speed,swets1961decision}. However, most previous work was focused on individual decision makers, and less is known about  groups of observers who make choices based on streams of evidence~\cite{gold2007neural,bose2017collective}. 

Each member of a social group often needs to choose between the same alternatives based on a combination of correlated and independent observations~\cite{kao2014decision}. 
For instance, when deciding whom to vote for, two individuals may see some of the same media coverage, but each may also read opinion pieces that the other does not~\cite{gerber2009does}. Conspecifics deciding 
where to forage are likely to rely on some of the same cues but can also learn from distinct experiences~\cite{valone1989group}. Traders may have access to private information but often track the same aggregate market indices and reports to decide what stocks to buy and sell, and the processes governing the valuation of distinct commodities are known to be correlated~\cite{mensi2013correlations}.
Thus, even in the absence of direct communication, the measurements individuals in a group use to make decisions are generally \emph{imperfectly} correlated.

Here, we assess the impact of such correlated measurements on the accuracy of individual decisions within groups of agents 
who do not share information. 
At first glance, it seems that such correlations may have no impact. How
can the accuracy of a decision by an isolated individual be influenced by someone else having access to the same information?
Indeed, when identical, rational agents make \emph{independent} observations the probability of a correct decision is unrelated to the order or the time at which the decision is made~\cite{karamched2020heterogeneity}: When measurements are independent, then so are the decisions of group members~\cite{moreno2010decision}.
This is no longer the case when a group of identical agents makes correlated measurements.
In this case early deciders tend to make decisions based on misleading observations, and their choices are less accurate than those of later deciders by as much as 20\%.  The order of a decision can therefore determines its accuracy.  However, any observer, although rational, believes that their decision is based on the same amount of evidence, and is therefore as accurate as that of anyone else. Yet, an outsider who observes the order in which decisions are made knows that early decisions are less likely to be correct than later ones. We analytically show why this is the case in tractable examples and provide an intuitive argument explaining why the same holds more generally.  Our analysis demonstrates why this difference in accuracy depends on how strongly evidence is correlated and on the size of the population. We conclude that pooling early decisions does not always help, but  weighting  individual decisions according to their order can produce better results.

\section{Model}
We consider a community of $N$ agents who accumulate evidence to decide between two states, or hypotheses, $H^+$ or $H^-$. Each agent accumulates evidence (observations) to decide between the two hypotheses. Agents are rational (Bayesian) and compute the probability that either hypothesis holds based on all evidence they accrue.  Each makes a decision once the log-likelihood ratio (LLR) of the conditional probabilities between the two hypotheses, given all the accumulated observations, crosses a predetermined threshold~\cite{Bogacz2006,Wald1948}. For simplicity, we assume that the observations the agents make are statistically identical and that they use the same decision policy. 
%We will discuss how these assumptions can be relaxed.
 
\paragraph{Independent evidence accumulation.} The problem of a single agent integrating evidence to decide between two options has been thoroughly studied~\cite{Bogacz2006, Gold2007,Ratcliff2008,Usher2001,veliz16,Wald1948}. In the simplest setting, an agent makes a sequence of noisy observations (measurements), $\xi_{1:t}$, with $\xi_i \in \Xi$ for $i \in \{1, \ldots, t\}$, where $\Xi \subset \mathbb{R}$. The observations, $\xi_i,$ are independent and  identically distributed, conditioned on the true state, $H \in \{\hp,\hmi\}$,
\[
P(\xi_{1:t} | H^{\pm} ) = \prod_{i=1}^t P(\xi_i | H^{\pm} ) =   \prod_{i=1}^t f_{\pm} ( \xi_i). 
\]
Here, the conditional probability of each measurement is given by the probability mass functions $f_{\pm}(\xi) := P(\xi | H^{\pm})$ when the conditional probability distributions are discrete, or by density functions when they are absolutely continuous. Observations, $\xi_i,$ are drawn from the same set, $\Xi,$ in either state $H^{\pm}$, and the two states are distinguished by the differences in the conditional probabilities of making certain measurements.  To simplify our analysis, we will sometimes assume that  only two observations are possible, $\Xi = \{\xi^+, \xi^-\}$.

To compute $P(H^{\pm}| \xi_{1:t})$, an ideal observer uses Bayes' rule.
For simplicity, we assume that the agent knows the measurement distributions, $f_{\pm}(\xi),$ and knows that both environmental states are equally likely, and hence uses a flat prior, $P(\hp) = P(\hmi) = 1/2$. The log-likelihood ratio (LLR) of the two states at time $t$ is then
\begin{align} \label{E:one_agent}
y_t & := \log \left( \frac{P( \hp | \xi_{1:t} )}{P( \hmi | \xi_{1:t} )} \right) = \sum_{s = 1}^{t} \LLR (  \xi_s) = y_{t - 1} + \LLR (  \xi_t),
%\sum_{s = 1}^{t} \log \left(  \frac{ P(  \xi_s  | \hp ) }{ P( \xi_s  | \hmi )} \right) = y_{t - 1} + \log \left( \frac{P(  \xi_t  | \hp ) }{P( \xi_t  | \hmi )} \right),
\end{align}
where $\LLR(\cdot) \equiv \log \frac{P(  \cdot  | \hp ) }{P( \cdot | \hmi )}$. We also 
refer to $y_t$ as the \emph{belief} of the agent at time $t$. The magnitude of the LLR can be viewed as the information an agent has gathered in support of a hypothesis, while its sign describes the proclivity of the agent. The flat prior implies $y_0 = \log \frac{P(H^+)}{P(H^-)} = \log \frac{1/2}{1/2}= 0$.  To understand the impact of common observations on decision probabilities in what follows, we constrain the beliefs, $y_t$, to the integer lattice by making further
assumptions about the distributions $f_+$ and $f_-$ .(See Appendix~\ref{A:lattice}.) 
%These assumptions simplify our arguments and make the calculations more transparent, but are not essential. 
%\WOcomment{I suggest that we do not discuss these further assumptions here.
%Rather, let us bring up assumptions on $\Xi$ and the two distributions on $\Xi$ when we discuss the simplified evidence model %(below).}

The optimality of the sequential probability ratio test~\cite{Wald1948} implies that an individual agent best manages speed and accuracy by waiting to decide until their belief reaches or crosses above (below) an upper (lower) threshold $\theta_+>0$ ($\theta_- <0$). Thus, an ideal agent continues making observations while $\thm < y_t < \thp $ and makes a decision after acquiring sufficient evidence,  choosing $\hp$ ($\hmi$) once $y_t \geq \thp$ ($y_t \leq \thm$). This fully defines an evidence accumulation model and decision policy for a single agent. We have analyzed a generalization of this model to social networks both small~\cite{Karamched20} and large~\cite{karamched2020heterogeneity}, where each agent accrues independent information according to Eq.~\eqref{E:one_agent} and shares their ongoing belief or decision state with some or all other agents in the group. These models of normative information-exchange based on neighbors' decisions build on previous work on normative confidence-weighting for majority rules~\cite{condorcet,nitzan1982optimal,boland1989majority,marshall2017individual}, locally-optimal Bayesian integration on sparse graphs~\cite{reina2022asynchrony}, the impact of common observations~\cite{moreno2010decision}, and non-normative decision sharing~\cite{Caginalp2017}. 
%Here, we assume that agents do not interact and make decisions without knowing if or which other agents in the group have decided. 
%Nonetheless, we will show that when agents collect a mixture of correlated and independent evidence, the accuracy of each decision depends on the timing of the given decision relative to those of others in the community.

\paragraph{Accumulation of correlated measurements.} 
To understand how correlated information impacts the accuracy of decisions in a community of $N$ agents, we assume that each agent acts in isolation. At each 
timestep, $t$, every agent, $i,$ makes an observation (measurement), $\xi^i_t \in \Xi$, and updates 
their private belief, $y^i_t,$ according to Eq.~\eqref{E:one_agent}. However, an individual agent does not know whether others have made decisions nor what those decisions were, in contrast to social network models studied in the past~\cite{Karamched20, karamched2020heterogeneity, Caginalp2017, condorcet, nitzan1982optimal, boland1989majority, marshall2017individual, reina2022asynchrony, olfati2006belief, banerjee1992, mossel2014opinion, ccelen2004observational}. This could be a model of a sample of voters, each of whom does not know the others or traders deciding to buy or sell without tipping their hand.

To model correlated measurements, we assume that with probability $c$ all agents make an identical observation on a timestep. An identical observation means that $\xi^i_t = \xi_t$ for all agents, $i = 1,...,N$, where $\xi_t$ is a single sample from the measurement distribution, $f_{\pm}(\xi)$. 
With probability $1-c$ agents make independent observations during a timestep, and the $N$ measurements, $\xi^i_t,$ are sampled independently from the distribution $f_{\pm}(\xi)$. 
This  is equivalent to having $N$ \emph{private}, independent sources of evidence, each accessible to a single agent, and one \emph{common} evidence source accessible to all agents (See the Discussion for less restrictive assumptions). 
%On each timestep, agents use the private sources or common source with probability $1-c$ and $c$, respectively. 
Therefore, the  belief of each agent evolves according to:
\begin{equation}
y^i_t  =  y^i_{t - 1} + (1-\chi_t) \cdot \LLR(\xi_t^i) + \chi_t \cdot \LLR(\xi_t),
\label{E:corr_eq}
\end{equation}
where $\chi_t$ are \emph{i.i.d.} Bernoulli random variables each with parameter $c$. 
%The quantity $c$  represents the strength of correlation in the observations: 
When $c = 1$ agents make only common observations, and when $c = 0$ agents make only independent observations. As $c$ increases from zero, each observation is more likely to be common, and the overall evidence becomes more correlated.
%The belief increments thus are identical with probability $c,$ and independent with probability $1-c$, and 
%Subsequently, the way in which observations are drawn on timestep $t$ is governed by the outcome of this Bernoulli random variable.
%Given the true state, $H^{\pm},$ and corresponding measurement distribution $f_s \in f_\pm$, we have
%$( \xi_t^1, ..., \xi_t^N ) \sim \prod_{i=1}^N f_s (\xi_t^i),$  if  $\alpha_t = 0,$ and
%$\xi_t^i \equiv \xi_t \sim f_s(\xi_t) $ for $i = 1, ..., N, $  if $\alpha_t = 1$.

Each agent makes observations until their belief, $y^i_t,$ reaches one of the thresholds, $\theta_{\pm}$, at which point they make the corresponding decision, $H^{\pm}$. 
For simplicity 
%henceforth, and since such an approach tends to best manage speed-accuracy tradeoffs in symmetric environments, 
we henceforth assume the thresholds are symmetric about zero, \emph{i.e.} $\theta_{\pm} = \pm \theta$, with $\theta >0$.  We denote the decision time of agent $i$ by $T_i$, and assume that 
decisions are immutable. Thus, decision times are uniquely defined, and only undecided agents continue to make observations. 

Importantly, agents \emph{do not} observe each others' decisions nor their current decision state (decided or undecided), in contrast with~\cite{Karamched20, karamched2020heterogeneity}.  Agents do not know whether an observation is common or private, and each uses the evidence they have collected 
to make the best possible decision based on their belief (LLR) given by Eq.~(\ref{E:corr_eq}). Hence, each agent acts as if they are alone.
%, and make a decision when their belief 
%crosses one of a pair of thresholds, $\{\theta_-, \theta_+\}.$ 

We ask how the accuracy of an agent's decision depends on the order in which the decision is made. %The first decider is the first agent whose belief reaches a threshold, $\pm \theta$, or, equivalently, the agent with the smallest decision time.
In particular, how accurate is the first decider?
If multiple agents make a decision at first-decision time, the `first' decider is chosen randomly with equal probability from that group.  The probability of a correct first decision then equals the probability that this first decider makes the correct choice, \emph{i.e.} that the belief of the first decider reaches the threshold, $\pm \theta,$ whose sign agrees with that of the true environmental state, $\hpm$. We briefly discuss other ways of defining a first decision in Appendix~\ref{app:1st}.

\paragraph{A simplified model of the evidence.} 
In some of our analyses and simulations, we assume that observers can make only two measurements, $\xi^+$ and $\xi^-$, \emph{i.e.} that both measurement distributions, $f_+$ and $f_-$, are concentrated on the same two values. We let $P(\xi^{\pm}| H^{\pm}) = p$ and $P(\xi^{\pm}| H^{\mp}) = q$   with $p + q = 1$ and $q< p.$ Thus, $p$ is the probability of making an observation consistent with the true hypothesis, while $q$ is the probability of an inconsistent observation. We assume $p/q = e$ hereafter so that the belief of each 
agent evolves on a lattice. This assumption simplifies the notation and analysis, but does not affect our conclusions.

\paragraph{Scaling limit of correlated evidence accumulation.} 
Computing decision accuracy and the distribution of decision times reduces to a first-passage problem~\cite{Bogacz2006}. 
Often, it is easier to solve such problems in the scaling limit, thus avoiding the combinatorial challenges common in discrete problems~\cite{redner2001}.
By invoking the Donsker Invariance Principle, in the limit of infinitely many infinitesimally informative measurements we obtain the macroscopic version of Eq.~\eqref{E:corr_eq}, often referred to as a {\em drift-diffusion equation}:
\begin{equation} dy_i = \pm \mu \, dt + \sqrt{ 2(1-c)\mu} \; dW_i + \sqrt{2c \mu} \; dW_c. 
\label{E:macroscopic}
\end{equation}
Here $y_i(t)$ is the limit of the LLR of agent $i$ and $\mu$ scales both the drift and diffusion terms. (See Appendix~\ref{macroscopic_derivation} for a derivation of Eq.~\eqref{E:macroscopic} and definition of $\mu$, which is proportional to the square of the signal-to-noise ratio of the sample distribution.)  The sign of the drift agrees with the sign of the environmental state, $H^{\pm}$. The Wiener processes $W_i(t)$ and $W_c(t)$ capture the variability of belief  increments due to independent and common observations, respectively.  Thus, the belief of each observer, $y_i(t)$, evolves according to a drift-diffusion model~\cite{Bogacz2006} that is coupled between observers~\cite{moreno2010decision}.

This model has been analyzed previously~\cite{moreno2010decision,shan2019family}, but we are not aware of a previous derivation from the normative model (See Discussion).
Eqs.~\eqref{E:corr_eq} and \eqref{E:macroscopic} both describe how beliefs evolve when agents collect both independent and common information. The derivation in Appendix~\ref{macroscopic_derivation} shows that the discrete and macroscopic models agree well when measurements carry little information and thresholds are large.

\section{Results}
%\begin{figure}[h!]
%\includegraphics[width = \textwidth]{fig2.pdf}
%\caption{Effect of correlated information on decider accuracy. (a) For $c\neq 0, 1$, the accuracy of the first decider dips below that of a single isolated decider (whose accuracy only depends on $\theta$). The minimum accuracy occurs for $c \approx 0.5$. (b) The later an agent in a network makes a decision, the more accurate they become. \textbf{Need an explanation of this panel.  I don't quite remember the details, but I remember the point was that the $N/2$th agent to decide displays an accuracy on par with the final decider for large $N$. I just don't remember what the dots mean.}  (c) In a community of two agents, $N = 2$, the average accuracy of the deciders equals their own assessed correct probability $(1 + \exp (-\theta))^{-1}$, as expected, but separating them by decision order reveals the accuracy of the first decider can be nearly 8\% lower than that of the second decider. (d) Decision times for agents in a network.   }
%\label{fig2}
%\end{figure}
\begin{figure}[h!]
\includegraphics[width = 12cm]{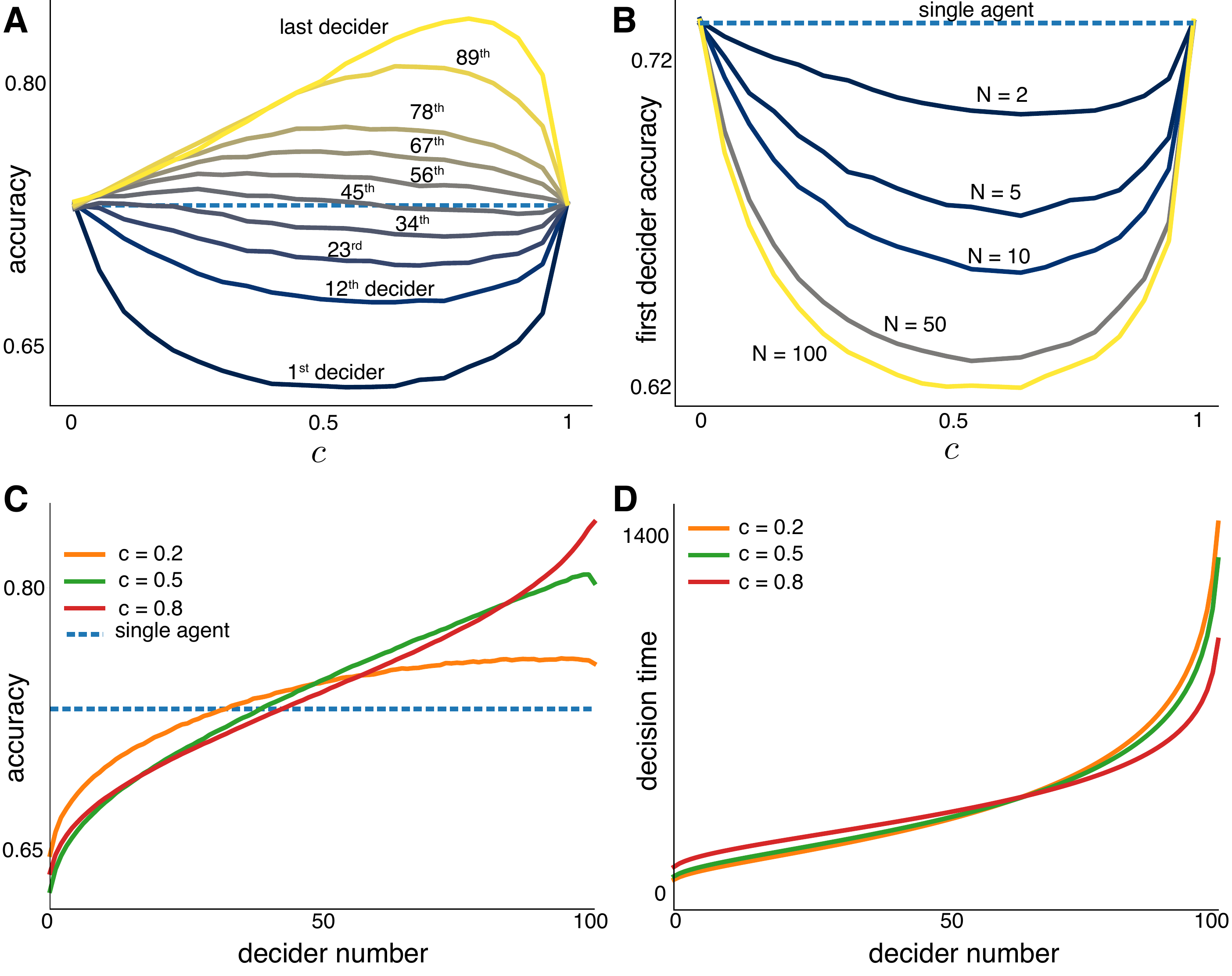}
\caption{Impact of the probability of making a common observation, $c$, on decider accuracy and timing. (A)~The probability of a correct decision increases with the order in which the decision is made. 
%The range of accuracies changes nonmonotonically with $c$. 
The average  accuracy computed over all deciders (dashed line) equals the accuracy of a randomly chosen agent, and is constant with $c$. Here, $N = 100$. (B) The accuracy of the first decider varies nonmonotonically with $c$, reaching a minimum at $c \approx 0.5$. As $N$ is increased, the minimum accuracy value decreases.  (C)~The accuracy of each of $N=100$ deciders increases with  decision order almost monotonically, so that the first (last) decider is less (more) accurate than a lone decider for $c \neq 0,1$. (D)~The time of the decision of $N=100$ agents as a function of order is approximately invariant for various  $c$ values, as the agents' beliefs become dominated by common evidence. Here $\theta=10$ and $p=e/(e+1)$.}
\label{fig2}
\end{figure}

We first asked how correlated evidence impacts the accuracy of decisions within a group of rational, identical agents.  Observers in our model do not know if their measurements are common or private, nor do they take into account the existence of other observers. Thus each of them makes decisions as if they were alone. The probability that a randomly selected agent in the group makes a correct choice does not depend on the number of other agents, nor on how strongly the evidence is correlated.  Thus it seems that correlations should not impact accuracy. But, surprisingly, for all $0< c < 1$, the  probability that the \emph{first}  decider in the group is correct is \emph{smaller} than the probability that a lone 
observer is correct. In particular, the first decider's accuracy reaches a minimum close to  $c = 0.5$ (Fig.~\ref{fig2}A). The probability that a decision is correct increases with the order in which the decision is made, so that later decisions tend to be more accurate than early ones (Fig.~\ref{fig2}B and C).
%\WOcomment{Are we sure that accuracy always increases as a function of decider order number?}
This result is the focus of our ensuing analysis. Decision times are more tightly distributed as common observations become more probable (Fig.~\ref{fig2}D), since observers' beliefs evolve more synchronously and thus cross the decision threshold at closer times.

That the probability of a decision being correct depends on its order is not intuitive: The beliefs of all agents in the community evolve according to identical stochastic processes. The agents set identical decision criteria (the thresholds $\pm \theta$) to gather what they believe is sufficient evidence in favor of one of the two choices. This decision threshold  determines the agents' perception of the probability that they will make a correct choice, ($1/(1+e^{-\theta}$)~\cite{wald1945,Bogacz2006}.
Since each agent decides as if they are alone, we originally expected that the accuracy of their decision would be determined only by the decision threshold, $\theta,$  and would thus be identical for all agents in the group.  Indeed,
the decision threshold determines the probability of a correct choice by an agent chosen at random at the outset of evidence integration.
%an agent at random the probability they 
%will make a correct decision is determined solely by this threshold, and is independent of the %correlation, $c$.
However, the first agent to make a decision is less likely to make a correct choice than all other agents in a group, and this probability 
decreases with the number of agents in the community (Fig.~\ref{fig2}B).
Further, decider accuracy increases almost monotonically with the order of the decision (Fig.~\ref{fig2}C).% except possibly for the last few deciders.
%\WOcomment{We need to check that we can defend these three claims.}
Thus, each agent perceives the same probability that their decision is correct, but someone observing the order in which decisions are made should trust later decisions more than early ones.   %Thus, 
%it becomes clear that conditioning on the order of an agent's decision introduces additional considerations into the calculation of their accuracy. In particular, 

The decreased accuracy of the first decider for $0 < c < 1$ relative to single-decider accuracy is not a trivial consequence of early deciders spending less time accumulating
evidence. If this were the case, the first decider would be less accurate than later ones when $c = 0$. But when observations are all independent, the probability of a correct decision is independent of the order in which the decision is made, and is determined  by the decision threshold.
Moreover, as $c$ increases from 0 to approximately 0.5, the average time to the first decision increases, but the average accuracy of this decision decreases. We next provide an explanation of this observation.

\subsection{An intuitive explanation for the decrease in first decision accuracy}
%\Zcom{Rewrite this part.}
 %We  start with an intuitive explanation of why correlated observations lead to less accurate first decisions.  We then provide and interpret an expression for the accuracy as a function of correlation, $c$, and community size, $N$.
 
Why do common observations lead to less accurate first decisions? The probability of a correct first decision when $c = 0$ or $c = 1$ equals the probability that the belief of a single observer evolving according to Eq.~\eqref{E:corr_eq} or Eq.~\eqref{E:macroscopic} crosses the correct boundary~\footnote{This probability can be calculated directly, or using first passage time methods~\cite{gardiner2009handbook}.
When the decision is triggered by the belief precisely meeting thresholds, it is given by inverting the log-likelihood ratio (LLR) to yield $(1+\exp(-\theta))^{-1}$~\cite{Bogacz2006}.}.
This is easy to understand: When $c = 0$, all agents make independent  observations, and the decision
accuracy of each agent, including the first, is determined by the decision threshold amplitude, $\theta$, alone. When $c = 1$, all agents receive the same evidence, so their beliefs evolve in unison, and the probability that they make a correct decision is thus the same as that of a single agent.

When $0 < c < 1$ the situation is different: The fact that other agents have yet to decide provides information that the first choice is less likely to be correct. At the time of the first decision, the undecided agents have likely made independent observations that counter the common observations that often contribute to the first decider's choice. Indeed, if these independent observations  agreed with the common evidence, the other agents would have more likely already made a decision too. For small $c$, little information is gained from common evidence, and not much independently gathered evidence is needed to counter it.
As $c$ increases, common evidence more often drives the first decision, so we expect that a substantial fraction of the independent evidence collected by an undecided agent will often counter the common evidence.
%Thus as $c$ increases, the amount of independent evidence counter to the first decision gathered by undecided agents also increases. 
However, when $c$ is large, most of the evidence is common, and fewer observations are independent, leaving less time for strong, contrary independent observations. Thus, at a critical value of $c$, the average total independent evidence obtained by undecided agents countering common observations reaches a maximum. The probability of a correct first decision is smallest at this critical value.  
%independent information to each information outweighs the asymmetry in survival probabilities in either the $H = H^+$ or $H = H^-$ states. 
In the next subsection, we make this argument more precise by showing that independent observations made by undecided agents that favor the correct decision are stronger when the first decider makes an
incorrect choice than in the opposite case.

\subsection{Reduction of the log-likelihood ratio of the first decider}

We next show mathematically why the decision of the first decider is less accurate than the decision of a randomly chosen agent.
By \textit{randomly chosen}, we refer to an agent chosen with equal probability among all agents in the group at the outset of evidence accumulation.  Equivalently, since all agents are identical we can compute  the probability that any given agent makes a correct decision.  % whereas the identity of the first decider is realization-dependent.
To do so, we write the log-likelihood ratio (LLR) corresponding to the conditional probability that the first decider makes the correct choice as a sum of two terms.   The first term corresponds to the
LLR of a randomly selected agent at decision time, while the second incorporates the condition that
the agent is the first decider. We show that the first term's magnitude equals that of the threshold, $\theta$, while the second term is negative for $0 < c < 1$. We conclude that the information obtained by undecided agents reduces the probability of a correct first decision. We begin by considering a pair of agents and obtain expressions for the sum of LLR terms in the case of beliefs evolving on a lattice. We then extend this calculation to an arbitrary number of agents and give intuition as to why agent accuracy is ordered by temporal decision order.

{\em Pair of agents in discrete time.} We randomly number the agents using indices $j = 1,2,$ and let $FD$ be the index of the first decider. Let $T_j$ be the time of the decision of agent $j$, and denote the decision of agent $j$  by $d_j \in \{ H^+, H^- \}$, so that $y_j(T_j) = \pm \theta$ and $|y_j(t)| < \theta$ when $0 \leq t < T_j$. Let $T = {\rm min}(T_1, T_2)$ denote the time of the first decision, so $T=T_1$ ($T = T_2$) if agent 1 (agent 2) is the first decider. Since the two hypotheses, $H^+$ and $H^-$, are equally likely,  we assume that the first decider chooses $H^+$ ($d_{FD} = H^+$), without loss of generality (WLOG).
We can therefore write the conditional probability $P^{\pm} (d_{FD} = H^+) := P(d_{FD} = H^+ | H^{\pm})$ as
\begin{align*}
    P^{\pm} (d_{FD} = H^+) &= \sum_{j=1}^2 P^{\pm} (d_j = H^+, FD = j) \\
    &= \sum_{j=1}^2 P^{\pm}(FD = j | d_j = H^+) P^{\pm} (d_j = H^+).
\end{align*}
Given the exchange symmetry between the two agents, $P^{\pm}(d_1 = H^+) = P^{\pm} (d_2 = H^+)$ and $P^{\pm}(FD = 1 | d_1 = H^+) = P^{\pm}(FD = 2 | d_2 = H^+) $, so that we can rewrite
\begin{align} \label{E:product}
P^{\pm}(d_{FD} = H^+) = 2 P^{\pm}(d_1 = H^+)P^{\pm}(FD = 1| d_1 = H^+).
\end{align}
The first term in the product is the $P^{\pm}$-probability that  a randomly chosen agent (here agent~1, WLOG) selects $H^+$. This term only depends on the measurements obtained by agent~1.  The second term is the $P^{\pm}$-probability that, conditioned on choosing $H^+$, agent 1 is also the first to decide. The second term  thus depends on the information gathered by the second agent. 

A randomly chosen agent, say $j=1$ as above, will be the first decider ($FD = 1$) if $T_1 < T_2$ and will be the first decider with probability $1/2$ if $T_{1} = T_{2}$. We expand the second term on the right side of Eq.~\eqref{E:product} assuming that $T_1, T_2 \in \mathbb{N}$ to obtain:
\begin{equation} \label{E:double}
\begin{aligned}
P^{\pm}(d_{FD} = H^+) = 2 P^{\pm}(d_1 = H^+) \sum_{t_1 \in \mathbb{N}} &\sum_{t_2 \in \mathbb{N}} P^{\pm} (FD = 1 | T_1 = t_{1}, T_2 = t_{2}, d_1 = H^+)
\\
&\qquad \quad {}\times P^{\pm}(T_1 = t_{1}, T_2 = t_{2} | d_1 = H^+).
\end{aligned}
\end{equation}
The first term in each product on the right side of Eq.~\eqref{E:double} reduces to 
\begin{align*}
    P^{\pm}(FD = 1| T_1=t_{1}, T_2=t_{2}, d_1 = H^+) = \left\{ \begin{array}{cc} 0, & t_1 > t_2, \\ 1/2, & t_1 = t_2, \\ 1, & t_1 < t_2. \end{array} \right.
\end{align*}
Thus, the inner sum over $t_2$ in Eq.~\eqref{E:double} gives terms corresponding to conditional complementary cumulative distributions of $T_2$ and conditional probabilities the agents decide simultaneously,
\begin{equation*}
\begin{aligned}
P^{\pm}(d_{FD} = H^+) = 2 P^{\pm}(d_1 = H^+) \sum_{t_1 \in \mathbb{N}} &\bigg[ \frac{1}{2} P^{\pm}(t_1 = T_{1} = T_2 | d_1 = H^+)
\\
&\qquad {}+ P^{\pm}(t_1 = T_{1} < T_2 | d_1 = H^+) \bigg].
\end{aligned}
\end{equation*}
%Thus, aside from the first decider's choice, we also observe the information provided by the indecision of the remaining agent. 
Using Eq.~\eqref{E:product} we can thus write the corresponding LLR of the first decider at the time of their decision as
\begin{align*}
    LLR(d_{FD} = H^+) = \log \frac{ P^+(d_{FD} = H^+)}{ P^-(d_{FD} = H^+) } =  LLR(d_1 = H^+) + LLR(FD = 1 | d_1 = H^+).
\end{align*}
The first term in this sum is the LLR of an individual observer, or, equivalently, a randomly chosen agent (taken here to be agent 1 WLOG), at the time of their decision. Hence, $LLR(d_1 = H^+) = \theta$.  The second term is given by
\begin{equation} \label{LLRdip}
\begin{aligned}
&LLR (FD = 1 | d_1 = H^+)
\\
&\qquad = \log \frac{\sum_{t_1 \in \mathbb{N}} \left[ \frac{1}{2} P^{+}(t_1 = T_{1} = T_2 | d_1 = H^+) + P^{+}(t_1 = T_{1} < T_2 | d_1 = H^+) \right]}{\sum_{t_1 \in \mathbb{N}} \left[ \frac{1}{2} P^{-}(t_1 = T_{1} = T_2 | d_1 = H^+) + P^{-}(t_1 = T_{1} < T_2 | d_1 = H^+) \right]}.
\end{aligned}
\end{equation}

Assume that agent 1 makes a wrong decision, \emph{i.e.} a decision inconsistent with the true hypothesis (here agent 1 does not necessarily make the first decision). In this case, both this agent's common and independent observations are likely to support the wrong decision.  But, by assumption,
any randomly sampled observation is more likely to be consistent with the true than the wrong hypothesis. Thus, the independent
observations of agent 2 are likely to point to the correct hypothesis and thus counter
the common observations supporting the incorrect decision of agent 1.  Hence, when agent 1 makes the wrong decision, agent 2 is more likely to have made conflicting observations: common observations consistent with the wrong hypothesis, and independent observations consistent with the correct one. As a result, agent 2 is more likely to decide after $T_1$, when the choice of agent 1  is wrong than when it is correct.  This argument shows that we expect
\begin{equation}
\label{E:key-inequality}
\begin{aligned}
&\sum_{t_1 \in \mathbb{N}} \left[ \frac{1}{2} P^{+}(t_1 = T_{1} = T_2 | d_1 = H^+) + P^{+}(t_1 = T_{1} < T_2 | d_1 = H^+) \right]
\\
&\qquad < \sum_{t_1 \in \mathbb{N}} \left[ \frac{1}{2} P^{-}(t_1 = T_{1} = T_2 | d_1 = H^+) + P^{-}(t_1 = T_{1} < T_2 | d_1 = H^+) \right]
\end{aligned}
\end{equation}
for $0 < c < 1$, so that Eq.~\eqref{LLRdip} implies $LLR (FD = 1 | d_1 = H^+) < 0$ for such values of $c$.
As a result, $LLR(d_{FD} = H^+) < LLR(d_i = H^+) = \theta$ for $i=1,2$ and $0<c<1$, so the first decider is less likely to make a correct choice than an agent chosen at random.

Moreover, as $c$ increases, so does the fraction of wrong common observations that can be countered by correct independent observations of agent 2.  This initially increases the likelihood that agent 2 remains undecided following incorrect decisions by agent 1.  But if $c$ is high, most observations are common, and agent 2  makes few independent observations.  Thus, as $c$ approaches 1 the agents' beliefs tend to evolve more synchronously, and the difference between the left and right sides of inequality~\eqref{E:key-inequality} decreases. This tension between the increase, with $c$, in the fraction of wrong common observations that are likely to be counteracted, and the decrease in the fraction of correct independent observations that can counteract them causes Eq.~\eqref{LLRdip} to achieve a minimum at an intermediate value, $0 < c < 1$.

%\Zcom{Any way to make a firmer argument that the above quantity will be negative?} We argue that when a randomly selected agent makes a decision consistent with the ground truth, it is less likely there will still be an undecided agent given some shared correlated evidence than in when the ground truth contradicts the selected agent's decision (Fig.~\ref{fig3}A,B). Thus, knowing an agent decided first reduces their probability of having made a correct choice.

\begin{figure}[t!]
\includegraphics[width = 11cm]{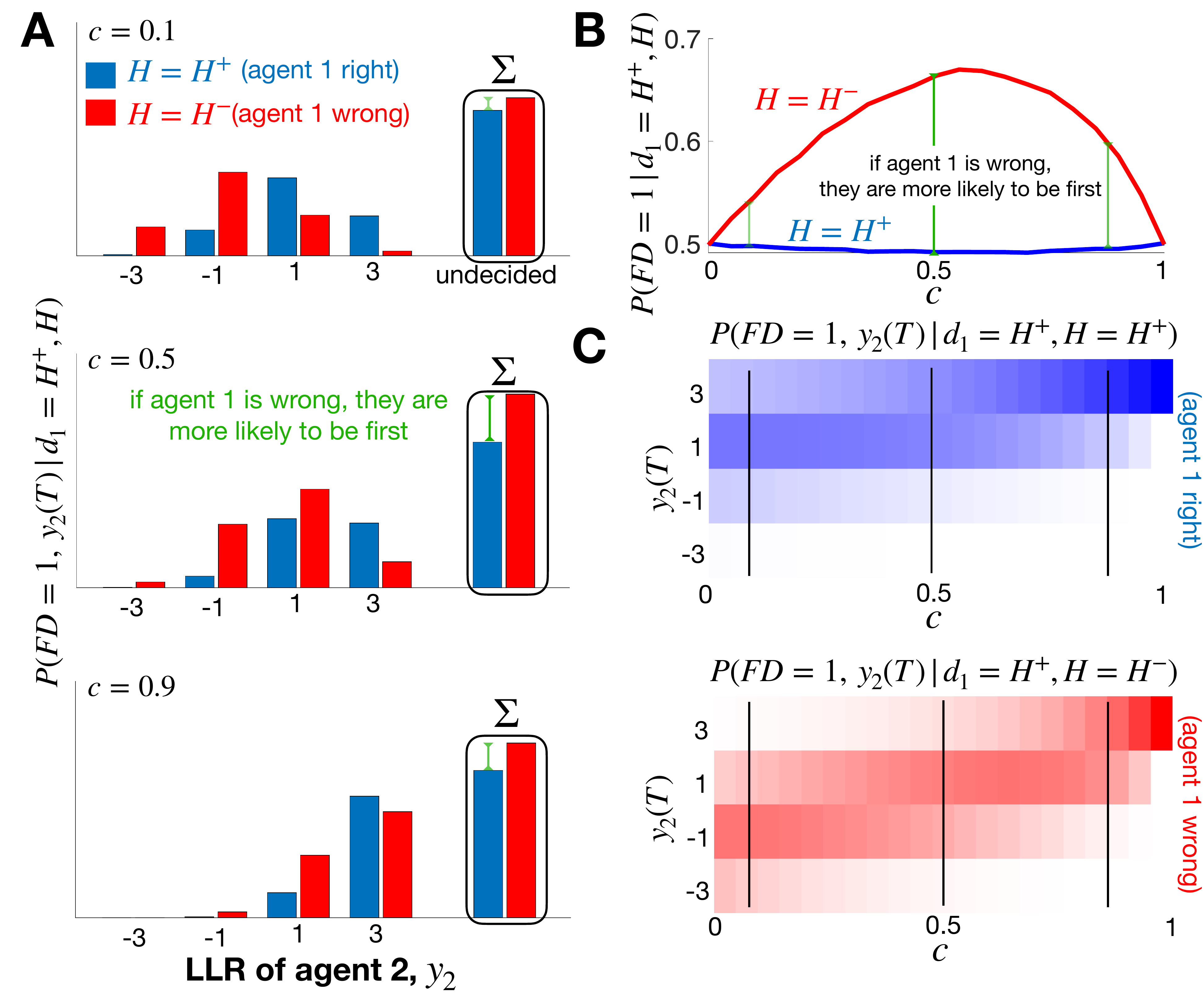}
\caption{When evidence is correlated a randomly selected agent is more likely to be the first decider if they are wrong.
(A)
Joint distribution of the probability  that agent 1 decides first ($FD=1$) and the belief of agent 2 at the time of the decision,  $y_2 = y_2(T)$, conditioned on agent 1 being right ($d_1 = H^+ = H$, blue) or wrong ($d_1 = H^+ \neq H^-$, red).
When $0<c<1$, the accuracy of the first decider is strictly below that of a randomly selected agent (here agent 1, WLOG) because of inequality~\eqref{E:key-inequality}.
When $c$ is small, $P^{+} (FD=1 | d_{1} = H^{+})$ nearly equals $P^{-} (FD=1 | d_{1} = H^{+})$ (difference indicated by green line), since the joint distributions are approximately reflections of one another, \emph{i.e.} $P^{+} (FD=1, y_{2}(T) | d_{1} = H^{+}) \approx P^{-} (FD=1, -y_{2}(T) | d_{1} = H^{+})$, with equality holding when $c = 0$.
As $c$ increases, the difference $P^{-} (FD=1 | d_{1} = H^{+}) - P^{+} (FD=1 | d_{1} = H^{+})$ first grows ($c=0.5$) and then shrinks ($c=0.9$) as both terms converge to $1/2$ as $c \to 1$. Note since each observation must increment an agent's belief $y_j$ by $\pm 1$, when $y_1(T) = \pm 3$, then $y_2(T)$ must take an odd value too.
(B)~The probability that agent 1 is the first decider (conditioned on $d_{1} = H^{+}$ and the environmental state) as  a function of $c$  peaks around $c=0.5$.
(C)
Colormap of the joint distributions from (A) as a function of $c$.}
\label{fig3}
\end{figure}

Numerical experiments support this explanation.
Fig.~\ref{fig3} illustrates the case of two agents, each with decision threshold magnitude, $\theta = 3$.
As our argument predicts, $P^{+} (FD=1 | d_{1} = H^{+}) < P^{-} (FD=1 | d_{1} = H^{+})$ for all $0 < c < 1$ (Fig.~\ref{fig3}B).
Further, the difference $P^{-} (FD=1 | d_{1} = H^{+}) - P^{+} (FD=1 | d_{1} = H^{+})$ first grows and then shrinks as $c$ increases, due mainly to the unimodalilty of the conditional probability that agent 1 decides first when their choice is wrong, $P^{-} (FD=1 | d_{1} = H^{+})$.
Looking at the joint conditional probabilities of $FD = 1$ and the belief of agent 2 at the time of the decision, $P^{+} (FD=1, y_2(T) | d_{1} = H^{+})$ and $P^{-} (FD=1, y_2(T) | d_{1} = H^{+})$  helps illuminate the situation.
Fig.~\ref{fig3}A shows these joint distributions for representative values of $c$ with $\theta = 3$.
The distribution of beliefs, $y_2,$ concentrates more on values $y_{2} (T) = \pm 1$ away from the thresholds, when $H = H^{-}$ than when $H = H^{+}$ for intermediate values of $c$ (Fig.~\ref{fig3}C).

Exact expressions for the probabilities in Eq.~\eqref{LLRdip} are unwieldy, but tractable solutions can be derived when two measurements are sufficient for belief magnitude to reach $\theta = 2$. We discuss this case next.

{\em Two agents with decision threshold parameter $\theta = 2$.} To obtain an explicit expression 
for the probability that the first decider is correct, we opt for a simplified model where only two measurement values ($\xi^{\pm}$) can be obtained  at discrete times, $\xi_t^i \in \{ -1, 1 \}, t = 1, 2, \ldots$, so  $f_{\pm}(\xi^{\pm}) = p$ and $f_{\pm}(\xi^{\mp}) = 1-p \equiv q$. Beliefs are restricted to the integer lattice by assuming $p = e /(e+1)$. When decision thresholds are set at $\pm \theta = \pm 2$, the belief of any undecided agent, $i,$ must equal $y_t^i = \pm 1$, at any odd time, and $y_t^i = 0$ at any even time~\footnote{Binarized evidence samples $\xi^{\pm}$ increment or decrement each agent's belief $y_t^i$ by one, so the sum of an even (odd) number of odd numbers, $\pm 1$, will be even (odd).}. Thus, the stochastic process governing the evidence accumulation of undecided agents resets to 0 (renews) every two timesteps, as long as an agent remains undecided.  If $T$ is the time of the first decision, then no agent could have reached either threshold before then, so we have  
\begin{align*}
    P(d_{FD} = H^{\pm}, T=t | T > t-2) = P(d_{FD} = H^{\pm}, T=2 )
\end{align*}
for all even $t > 0$, since if $T > t-2$ then at time $t-2$ both agents must have been undecided and thus had beliefs $y_{t-2}^i = 0$.
%This renewal property eases the calculation of the constituent functions within the LLR we computed above for the first decider.

We now enumerate and sum the probabilities of all cases in which agent 1 (not necessarily the first decider) makes decision  $d_1 = H^+$ under either condition, $H = H^{\pm}$. To begin, there are four ways for the two agents to make a decision on the same timestep: If $d_1 = H^+$, then the second agent can make the same decision 
simultaneously ($d_2 = H^+$) if the second agent made zero, one, or two independent measurements, or the second agent can make the opposite choice simultaneously ($d_2 = H^-$),  if they made two independent measurements. Therefore,
\begin{align*}
    P^+(T_1 = T_2 = t | d_1 = H^+, T>t-2) &= c^2 + 2c(1-c)p + (1-c)^2(p^2 + q^2), \\
    P^-(T_1 = T_2 = t | d_1 = H^+, T> t-2) &= c^2 + 2c(1-c)q + (1-c)^2(q^2 + p^2). 
\end{align*}
The second agent may remain undecided at the time of the first agent's decision if they made one independent measurement that conflicts with the first agent's decision, or two independent measurements that conflict with each other:
\begin{equation}
\begin{split} \label{E:indecision}
    P^+(T_1 =t < T_2 | d_1 = H^+, T>t-2) &= 2c(1-c)q + 2(1-c)^2pq, \\
    P^-(T_1 =t< T_2 | d_1 = H^+, T>t-2) &= 2c(1-c)p + 2(1-c)^2pq.
    \end{split}
\end{equation}
%The probability of the first decision time exceeds a given even value $t$, is the probability that the belief of both agents returned to zero at $t$ and at all preceding even times. Due to the renewal property, we can compute this for $t=2$ and the powers for all even times thereafter:
%\begin{align*}
%    P^{\pm} (T>t = 2m) &= P(T>2)^m, \\ 
%    P(T>2) &= 2 c^2pq + 2c(1-c)(pq^2+p^2q)+4(1-c)^2p^2q^2.
%\end{align*}
Now let $m \in \mathbb{N}$ and $t_{1} = 2m$.
Referring to the sums in Eq.~\eqref{LLRdip}, we have
\begin{align*}
&P^{\pm} (t_{1} = T_{1} = T_{2} | d_{1} = H^{+})
\\
&\qquad =
P^{\pm} (t_{1} = T_{1} = T_{2} | d_{1} = H^{+}, T > t_{1} - 2) P^{\pm} (T > t_{1} - 2 | d_{1} = H^{+})
\\
&\qquad = P^{\pm} (t_{1} = T_{1} = T_{2} | d_{1} = H^{+}, T > t_{1} - 2) P (T > t_{1} - 2)
\\
&\qquad = P^{\pm} (t_{1} = T_{1} = T_{2} | d_{1} = H^{+}, T > t_{1} - 2) [P(T>2)]^{m-1}.
\end{align*}
A similar calculation gives
\begin{equation*}
P^{\pm} (t_{1} = T_{1} < T_{2} | d_{1} = H^{+}) = P^{\pm} (t_{1} = T_{1} < T_{2} | d_{1} = H^{+}, T > t_{1} - 2) [P(T>2)]^{m-1}.
\end{equation*}

We can factor the common terms out of the sums in Eq.~(\ref{LLRdip}) and cancel sums over the factors of $P(T>2)^{m-1}$ in the numerator and denominator to obtain an explicit form of Eq.~\eqref{LLRdip},
\begin{align}
    LLR(FD=1|d_1 = H^+) = \log \frac{\left[ c^2 + 2c(1-c)(1+q) + (1-c)^2 (1+2pq) \right] }{ \left[ c^2 + 2c(1-c)(1+p) + (1-c)^2 (1+2pq) \right]  }. \label{LLRdip_h2}
\end{align}
The numerator and the denominator in this expression differ only in the middle terms, $2c(1-c)(1+p) > 2c(1-c)(1+q)$ for $0<c<1$. 
Eq.~\eqref{E:indecision} shows that this term corresponds to the probability that agent 2 makes an independent observation that counters the common observation of the two agents, in agreement with our general explanation.  As discussed previously, this is more likely when the decision of the first agent (and the common measurement) is wrong.  

%revealing the information obtained from the unchosen agent's indecision is due to the difference in the middle term, arising realizations in which they receive a single independent and contrary observation. All other terms are the same in the numerator and denominator, and so since $p>q$, the formula in Eq.~(\ref{LLRdip_h2}) will be negative. This implies that knowledge of a remaining agent's indecision reduces the confidence one should have in the first decision.

{\em Continuum case.}
Our results for the discrete model extend to agents with continuously evolving beliefs, obtained in the limit of many weak observations (see Appendix~\ref{macroscopic_derivation}).
The beliefs, $y_i(t)$,  evolve according to Eq.~(\ref{E:macroscopic}) and decisions are determined  by the threshold each belief trajectory reaches first. Each agent's decision time is a positive real number,  $T_j \in (0, \infty)$  for $j=1,...,N$, so $\mathbf{T} = (T_{1}, \ldots , T_{N}) \in (0, \infty)^N \equiv \mathbb{R}_+^N$. For finite $N$ and $c<1$, the probability that two agents decide at the same time is zero. This last observation simplifies our arguments, since we do not need to account for simultaneous decisions, as we do in the discrete model.

\begin{figure}[t!]
\includegraphics[width = \textwidth]{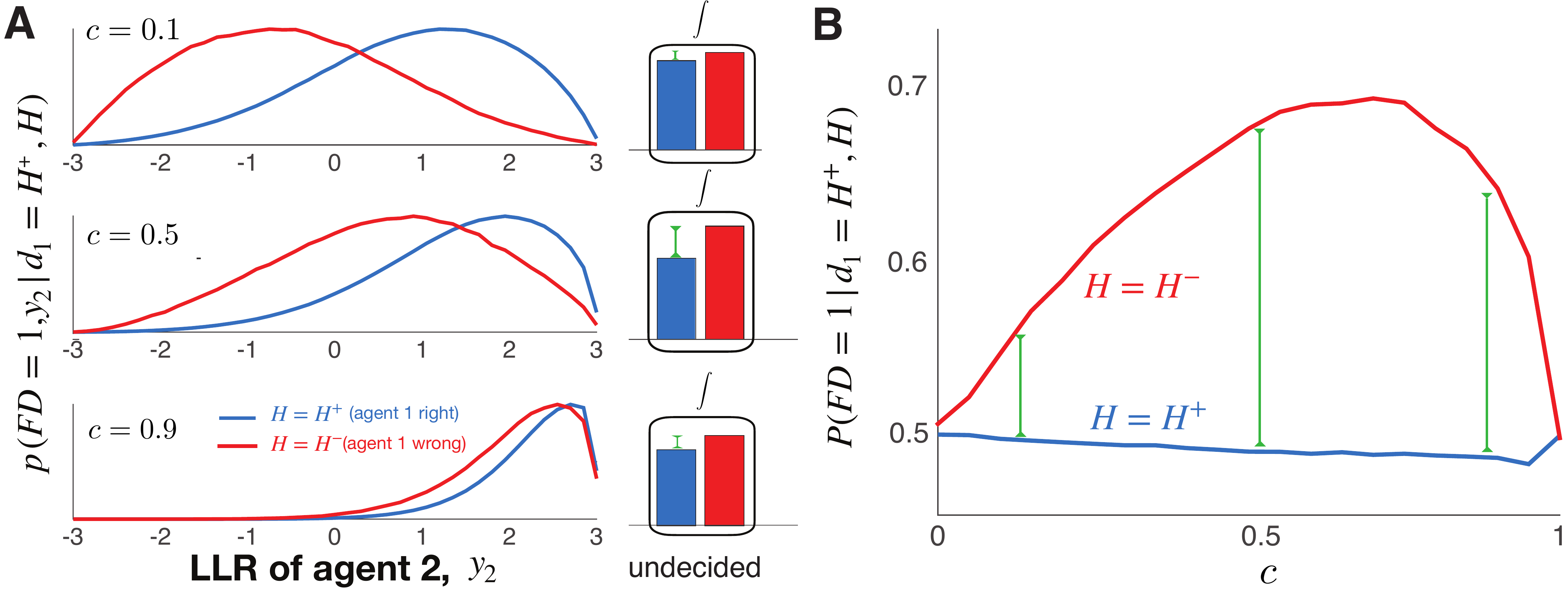}
\caption{When beliefs evolve continuously and evidence is correlated, a random agent is again more likely to decide first if they are wrong.
(A)~As in the discrete model, the densities $p(FD=1, y_{2} | d_{1} = H^{+}, H)$ are nearly reflections of one another for small $c$.
By marginalizing over the distribution of beliefs, $y_2$, we can obtain the difference $P^{-} (FD=1 | d_{1} = H^{+}) - P^{+} (FD=1 | d_{1} = H^{+})$. This difference is small when $c$ is small (red bar minus blue bar).
As $c$ increases, this difference first increases and then decreases, the latter because each term in the difference coverges to $1/2$ as $c \to 1$.
(B)~The unimodal dependence of first-decider accuracy as $c$ increases is again due to $P^{-} (FD=1 | d_{1} = H^{+})$ obtaining a maximum  around $c = 0.5$. 
%due to the maximal possible increment of commmon observations being countered by independent observations at intermediate values of $c$ (red curve), while $P^{+} (FD=1 | d_{1} = H^{+})$ is essentially insensitive to $c$ (blue curve).
}
\label{fig4}
\end{figure}
By marginalizing over all agents and decision times, we obtain the counterpart of Eq.~\eqref{E:double},
\begin{align*}
    P^{\pm}(d_{FD} = H^+ ) = N \cdot P^{\pm} (d_1 = H^+) \int_{\mathbb{R}_+^N} P^{\pm}(FD = 1 | \mathbf{T} = \bm{t}, d_1 = H^+) g^{\pm}(\bm{t} | d_1 = H^+) \, d \bm{t}.
\end{align*}
Here $g^{\pm} (\cdot | d_{1} = H^{+})$ is the conditional probability density function for the time of the first decision, $\mathbf{T}$, conditioned on the state,  $H = H^+$ or $H = H^-$, and on the decision $d_{1} = H^{+}$.
We have $ P^{\pm}(FD = 1 | \mathbf{T} = \bm{t}, d_1 = H^+) = 1$ if $t_{1} = \min_{1 \leqslant i \leqslant N} t_{i}$, otherwise this quantity is zero.
This simplifies the multi-dimensional integral in the preceding expression to an integral over the $t_1$-axis,
\begin{align*}
    P^{\pm} (d_{FD} = H^+) = N \cdot P^{\pm}(d_1 = H^+) \int_{\mathbb{R}_+} G^{\pm} (t_{1} | d_{1} = H^{+}) \, d t_{1}.
\end{align*}
Here
\begin{equation*}
G^{\pm} (t_{1} | d_{1} = H^{+}) = \int_{(t_{1}, \infty )^{N-1}} g^{\pm} (t_{1}, t_{2}, \ldots , t_{N} | d_{1} = H^{+}) \, d t_{2} \cdots d t_{N},
\end{equation*}
is the probability density of the time of the decision of agent 1, given that agent 1 chooses $H^+$, and the other agents are undecided.
We can thus write
\begin{align*}
LLR (d_{FD} = H^{+}) = \theta + \log \frac{\int_{\mathbb{R}_+} G^{+} (t_{1} | d_{1} = H^{+}) \, d t_{1}}{\int_{\mathbb{R}_+} G^{-} (t_{1} | d_{1} = H^{+}) \, d t_{1}}.
\end{align*}
As in the discrete case, the second term is the log of the ratio of conditional probabilities that all other agents remain undecided at the time agent 1 chooses $H^+$. 
%As we show in Fig.~\ref{fig3}, this term is negative for all $0< c < 1$ for the same reasons as the discrete case.
%\Kcom{This ends a bit flat, but that may be OK.}

When $N = 2$  the nonmonotonicity of the first decider's accuracy in $c$ is due to the increasing opportunity for conflict between the second agent's independent observations and common observations shared by both agents (agents 1 and 2), but a decreasing probability of independent observations, as $c$ increases.
The densities $\frac{d}{dz} P(FD=1, y_{2}(T) \leqslant z | d_{1} = H^{+}, H)$ are nearly reflections of one another for small $c$ (Fig.~\ref{fig4}A, top left).
Integrating over $z$, the difference $P^{-} (FD=1 | d_{1} = H^{+}) - P^{+} (FD=1 | d_{1} = H^{+})$ is small when $c$ is small (red bar minus blue bar, Fig.~\ref{fig4}A, top center).
For intermediate values of $c$, the distribution of beliefs of agent 2 is pulled away from the correct threshold when the randomly selected agent makes the wrong choice, due to common observations, causing $P^{-} (FD=1 | d_{1} = H^{+}) - P^{+} (FD=1 | d_{1} = H^{+})$ to reach a maximum within the intermediate $c$ range.
When $c$ is close to $1$, both $P^{-} (FD=1 | d_{1} = H^{+})$ and $P^{+} (FD=1 | d_{1} = H^{+})$ converge to $1/2$, so the difference converges to zero.
Fig.~\ref{fig4}B shows that the unimodal response of first-decider accuracy as $c$ increases occurs because the probability $P^{-} (FD=1 | d_{1} = H^{+})$ of an incorrect agent deciding first increases for small $c$ and then decreases in $c0$ (red curve), 
%since the largest incorrect correlated increments can be overcome at intermediate $c$ (red curve), 
while $P^{+} (FD=1 | d_{1} = H^{+})$ is approximately insensitive to $c$ (blue curve). 

{\em Extending the analysis of the discrete model to more than two agents.}
When there are more than two agents, $N>2$, numerical 
simulations show that accuracy of the first decider can dip even lower (See Fig.~\ref{fig2}B). To explain 
this more general observation, we extend our two-agent analysis. We denote the decision of agent $j \in \{1, \ldots, N\}$ by $d_j$, and the corresponding decision time by $T_j$. The probability that the first decider chooses $H^+$ conditioned on  the true  state is given by
\begin{align*}
    P^{\pm}(d_{FD} = H^+) = \sum_{j=1}^N P^{\pm}(FD = j | d_j = H^+) P^{\pm}(d_j = H^+). 
\end{align*}
Exchange symmetry again implies that  $P^{\pm}(d_j = H^+) = P^{\pm}(d_1 = H^+)$ and $P^{\pm}(FD=j | d_j = H^+) = P^{\pm}(FD=1 | d_1 = H^+)$ for all $j \in \{1, \ldots, N\},$ and so
\begin{align*}
    P^{\pm}(d_{FD} = H^+) = N \cdot P^{\pm}(d_1 = H^+) P^{\pm} (FD = 1 | d_1 = H^+).
\end{align*}
Let $\mathbf{T} = (T_1, ..., T_N) \in \mathbb{N}^N$ be the vector of decision times and let $T = \min_{j} \{T_j\}$ be the time of the first decision.  Then 
\begin{equation}
\label{E:many}
\begin{aligned}
P^{\pm}(d_{FD} = H^+) &= N \cdot P^{\pm}(d_1 = H^+)
\\
&\qquad \times \sum_{\bm{t} \in \mathbb{N}^N} P^{\pm} (FD = 1 | \mathbf{T} = \bm{t}, d_1 = H^+) P^{\pm} (\mathbf{T} = \bm{t} | d_1 = H^+),
\end{aligned}
\end{equation}
where the first term in the sum vanishes if $t_1 > \min_{1 \leqslant j \leqslant N} t_{j}$. On the other hand, if $t_1 = \min_{1 \leqslant j \leqslant N} t_{j}$, the conditional probability that agent 1 is chosen as the first decider depends on the number of indices $j$ for which $t_{j} = t_{1}$, \emph{i.e.} the number of agents who simultaneously make the first decision.
Let $n_{FD}(\bm{t})$ denote the number of these first deciders.
Overall, we have
\begin{align*}
P^{\pm}(FD = 1 | \mathbf{T} = \bm{t}, d_1 = H^+) = \left\{ \begin{array}{cc} 0,   & t_1 > \min_{1 \leqslant j \leqslant N} t_{j}, \\ 1/ n_{FD}(\bm{t}), \ \quad & t_1 = \min_{1 \leqslant j \leqslant N} t_{j}. \end{array} \right.
\end{align*}
Thus, we can turn the second term within the sum from Eq.~\eqref{E:many} into an additional sum over the count of agents deciding at the first decision time:
\begin{align*}
    P^{\pm}(d_{FD}= H^+) = N \cdot P^{\pm}(d_1 = H^+) \sum_{t_1 \in \mathbb{N}} \sum_{k=1}^N \frac{1}{k} \cdot P^{\pm}(t_{1} = T_1 = T, n_{FD}(\mathbf{T}) = k | d_1 = H^+).
\end{align*}

As before, we write the LLR as a sum of two terms, one given by the LLR of a randomly selected agent (agent 1) choosing $H^+$, 
$LLR(d_1 = H^+) = \log (P^+(d_1 = H^+))/P^-(d_1 = H^+) = \theta$, and a second term involving conditional probabilities that the randomly selected agent is the first decider,
\begin{align*}
    LLR(d_{FD} = H^+) = LLR(d_1 = H^+) + LLR (FD = 1 | d_1 = H^+),
\end{align*}
where 
\begin{align}
LLR(FD = 1 | d_1 = H^+) = \log \frac{\sum_{t_1 \in \mathbb{N}} \sum_{k=1}^N \frac{1}{k} \cdot P^{+}(t_1 = T_{1} = T, n_{FD}(\mathbf{T}) = k | d_1 = H^+)}{\sum_{t_1 \in \mathbb{N}} \sum_{k=1}^N \frac{1}{k} \cdot P^{-}(t_{1} = T_1 = T, n_{FD}(\mathbf{T}) = k | d_1 = H^+)}.    \label{LLRdip_N}
\end{align}
%Again, we expect that when a randomly chosen agent's decision conflicts with the ground truth, it is more likely they will be the first decider. As in the case of a pair of agents, the difference between the term in the numerator and denominator arises primarily due to cases in which erroneous correlated evidence pushes all agents toward the wrong threshold and independent observations push most but not agents back toward the true threshold. One or a few agents will make the wrong choice, and more likely be chosen as first deciders in these cases. 

This term has the same form as in the case of two agents, and is negative for $0 < c < 1$ for the same reason: Common observations are likely to be in agreement with the decision of the first decider.  However, when the first decider is wrong, independent observations of the other observers are more likely to point in the direction opposite of the first decision than when the first decision is correct.  Thus, the first decider is less likely to be correct than a randomly selected agent when $0 < c < 1$, in agreement with simulation results. Moreover, the difference between the numerator and denominator grows with the number of agents, reflecting the additional information provided by having even more undecided agents (Fig.~\ref{fig2}B).
Other agents will make observations countering
the first decision when it is incorrect, and consistent with it when it is correct.

{\em Accuracy depends almost monotonically on temporal decision order.}
Fig.~\ref{fig2}A illustrates that decider accuracy increases almost monotonically with position in the temporal decision order. We hope to explore exactly in what situations non-monotonicity emerges in future work. %, except, possibly, for the last few deciders.
To explain this relation, consider computing the LLR at the time of decision for the second decider. We proceed with the same computation as above, but, in this case, fewer agents remain undecided.
On average, we thus expect the number of independent observations by undecided observers inconsistent with a wrong $m^{\mathrm{th}}$ choice, and the number of observations consistent with a correct $m^{\mathrm{th}}$ choice to diminish with the order, $m$, of the decision.  We thus expect the ratio between the numerator and denominator in Eq.~\eqref{LLRdip_N} to be greater than that in the case of the first decider. Repeating the argument recursively for subsequent agents suggests that accuracy increases with decision order.  Such an argument provides intuition, though we admit it is incomplete as the computation would have to take into account the observations of both agents that make a decision before and after a given agent. 

Why does the accuracy of deciders that decide near the end of the temporal decision order exceed the accuracy determined by the threshold $\theta$? The average accuracy of all agents is determined by the threshold $\theta$. 
As such, accuracy that falls below that which is determined by $\theta$ must be counteracted by agent accuracy that exceeds that value.
%\WOcomment{Shall we move this ordering of accuracies material to the discussion?} \Kcom{I think it is fine here.}

%Randomly selecting two agents out of $N>2$ and applying the formula Eq.~(\ref{LLRdip}) will generally imply the first agent to decide is less accurate (assuming this additional term is negative). Repeating this process exhaustively then implies that agents' accuracies are ordered according to their decision time order. A similar argument can be applied using random subsets of any size and using Eq.~(\ref{LLRdip_N}).

%\Kcom{I don't follow this.  I think the argument is something like this:  Let's take the second agent, and go through the same computation as above. But then there are fewer agents that need to remain undecided, so that twe expect the difference between the numerator and denominator in Eq.~\eqref{LLRdip_N} to be smaller than in the case of the first agent. Now repeat recursively until the last agent. This does not answer why the second half of the agents now needs to be more accurate than the first.  However, we could appeal to the argument that on average they have to have accuracy determined by the threshold $\theta$, so some must be more accurate than $\theta$.}

\subsection{Pooling over early deciders does not rescue accuracy}

\begin{figure}[t!]
\includegraphics[width = \textwidth]{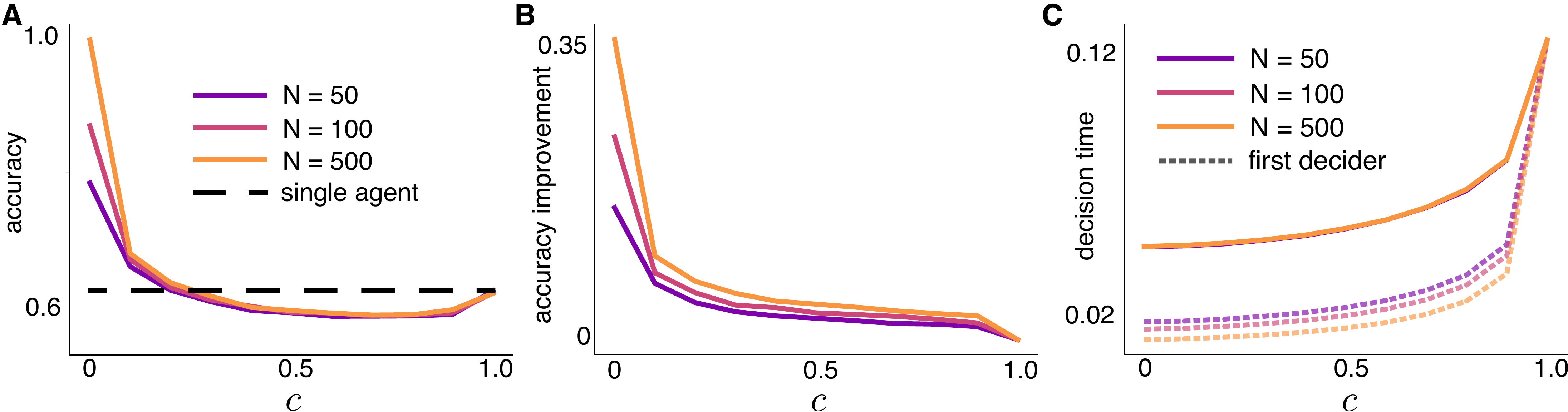}
\caption{Pooling choices of early deciders using a majority rule only mildly improves accuracy compared to the first decision when evidence is correlated.
(A)~The group's decision is determined by the majority of the first $N_{\rm pool}$ deciders.
 For different population sizes, $N$, the accuracy of the group decision at first decreases as $c$ is increased, and can be lower than the accuracy of a single decider in isolation (dashed line). Here  $N_{\rm pool} = 0.2 \cdot N$.
(B)~Improvement in the accuracy obtained by pooling the first $N_{\rm pool} = 0.2 \cdot N$ decisions compared with the accuracy of the first decision drops substantially even for small values of $c$, and is nearly independent of $N$.
(C)~The mean time at which the last decider in the pool makes a decision increases with $c$ (solid curves).
Dashed curves give the mean time of the first decision.
}
\label{fig5}
\end{figure}

The ``wisdom of crowds'' is the idea that collective decision by a group of people is more likely to be correct than the decision of any single member of the group~\cite{condorcet,de2014essai,galton1907vox}.  The group's accuracy can be improved when individuals exchange information preceding their final decisions or when the group decision is determined by the majority of individual choices~\cite{condorcet,karamched2020heterogeneity, reina2022asynchrony, conradt2005, surowiecki2005wisdom}. However, this improvement can be diminished, and individuals can even outperform crowds when biases in individual decisions are not accounted for when forming the group decision~\cite{simmons2011intuitive,lo2022wisdom}. Here we show that when agents receive even modestly correlated information, group decisions obtained by pooling the choices a group of early deciders can still be less likely to be right than the decision of a randomly selected agent. Moreover, such pooling only modestly improves on the accuracy of the first decider (Fig.~\ref{fig5}A,B). The additional time required to obtain these additional opinions is not negligible, and is roughly independent of the population size, $N$ (Fig.~\ref{fig5}C). 
Hence, even weak correlations in the evidence can impact the accuracy of group decisions.

\section{Discussion}

Humans and other animals integrate evidence to make decisions. Often members of a group or community are faced with the same choices and will use evidence that is available to all of them to decide between a common set of options~\cite{seeley1991collective,pratt2002quorum,Couzin2009}. We have shown that when some observations are made in common, even when no 
social information is exchanged the first individual to decide makes the least accurate decision.
The accuracy of subsequent decisions increases in the order in which they are made, with few exceptions. 

We have focused on agents deciding between two options, so that response accuracy can be computed as exit probabilities of populations of univariate stochastic processes driven by common and independent noise~\cite{gardiner2009handbook}. The accuracy of the first agent to make a decision depends non-monotonically on the probability $c$ of making a common measurement. When the accuracy of the first decision is at a minimum, roughly half the observations are common. The remaining independent observations allow the agent's beliefs to diverge, leading the first agent to often choose differently than later deciders.

We made the simplifying assumption that all agents either jointly make a common observation or all make private observations on each timestep.  This requires a coordinated measurement process, which is counter to our assumption that agents do not share social information. We could relax this assumption and allow agents to each independently make measurements from two sources, one common to the group and one available only to the agent. With two agents this model is equivalent to the model we analyzed. 
More generally, different subsets of agents could have access to 
separate sources of shared information, rather than a single common source available to the entire community.
The analysis of these cases becomes more cumbersome, but we expect that our general conclusions will hold.

Such generalizations are related to a subtle point: If agents share their decisions, but not their measurements, then the fact that no decision has been announced up to a time $t$ is itself informative. Suppose evidence is correlated, and the belief of a given agent is the first to reach threshold. That agent knows that no other decisions have been made. Indeed with imperfectly correlated information, the absence of a decision can reveal information about the evidence other agents have gathered, and a rational agent would
take this into account. They may be using the same reasoning we used to explain the 
observed drop in accuracy, thinking: ``I have gathered sufficient information to make a choice, and everyone else is undecided. They are likely using some of the same information as me, but are not yet convinced. Perhaps I should distrust some of the information I have obtained.''
We have shown previously that similar reasoning can lead to intricate social information 
exchanges~\cite{Karamched20}. However, humans frequently exhibit 
correlation neglect~\cite{enke19}. If observers assume that information is uncorrelated, then the model we described here may be applicable even when they observe each other's decisions.

We have assumed that the agents in the population are identical.  If agents have different decision thresholds, early decisions tend to be driven by less evidence~\cite{Bogacz2006}, generating a decrease in accuracy unrelated to the effect of common observations. Correlated evidence could exacerbate this decrease in accuracy. However, if agents have access to information of different quality, early deciders tend to be those with access
to the best information~\cite{reina2022asynchrony}. In this case early decisions can be more accurate than later ones.  We expect correlated evidence to still impact the accuracy of the first decision, but the specifics would depend on the quality of common and private evidence.

Except for limiting cases, we found it quite cumbersome to obtain analytical expressions for the accuracy of the first decider and other statistics of the agents' decisions. However, prior work has shown that the correlated drift diffusion model generated in the macroscopic limit can be solved explicitly using method of images solutions for specific threshold values~\cite{shan2019family}. In our case, thresholds are always square for $N=2$ or cubes/hypercubes for $N>2$, but it is possible a method of images approach may still apply. 

 Like other mathematical models of cognition, our model only roughly approximates decision-making process used by humans and animals. Despite its limitations, we believe that our analysis offers insights that are independent of the exact way in which evidence is integrated and decisions formed in groups. The essential idea behind the effect we describe is that common observations drive the beliefs of individuals in the community in the same direction. If those common observations are misleading, it takes time for private evidence to counter their effect. When deciders use a substantial fraction of common observations to make their decisions, early decisions are most likely consistent with common observations. Thus, if common observations are right (wrong), the first decision tends to be as well. First decisions thus tend to be based predominantly on common evidence, which offers less information than what is implied by the decision threshold.
We expect that the resulting asymmetric weight of common evidence in determining the first decision leads to similar effects more generally, \emph{e.g.}, when the population is heterogeneous, faced with more than two choices, or when observations are made asynchronously.   
Social information exchange would lead to more subtle effects, modulating the impact of  common measurements.  We
have thus described a general mechanism that can affect group decision-making, with implications that transcend specific scenarios. The insights we provided can describe decision-making processes across a range of contexts and could be used to lead to more effective individual and group choices.

%\Kcom{
%\begin{enumerate}
    %\item Discuss alternative model for common observations. Point to appendix for some details.
  %  \item Discuss what happens if agents can observe each other's choices
%\end{enumerate}
%}

\appendix

\section{Beliefs evolving on the integer lattice} \label{A:lattice}
In the simplest case we can assume that an observer can make only two measurements, $\xi^+$ and $\xi^-$. We let $P(\xi^{\pm}| H^{\pm}) = p$ and $P(\xi^{\pm}| H^{\mp}) = q$   with $p + q = 1$ and $q< p.$  Assuming $q = p/e$, gives $p + p/e = 1$ so that $p = e/(1+e),$ $q=1/(1+e)$, and  hence,  $\text{LLR}(\xi^{\pm}) = \pm 1.$  In particular, when  $q = p/a$ then the information provided by observation $\xi_{\pm}$ equals $\pm \ln a.$
As a result, the belief of each agent, $y_i$, lies on a lattice defined by $\{ n \ln a \}_{n \in \mathbb{Z}}$, and we can use the mapping $ n \rightarrow n \ln a$ or a 
logarithm in base $a$ to place beliefs on an integer lattice.
In the double limit of infinitesimal evidence, $\ln a$, ($a \to 1^+$), and infinitesimal time between observations, we recover a continuous model as outlined in the next section.  

\section{Derivation of the scaling limit}
\label{macroscopic_derivation}

Let $f_+(\xi)$ be a probability distribution of  observations, $\xi$, over an arbitrary set $\Xi$ obtained in state $H^+$, and $f_-(\xi)$ the probability distribution of observations over that same set $\Xi$  in state $H^-$. Note if the sets of observations for either state differ, then there will be infinitely informative observations which, when observed, would immediately make an agent certain of the state. However, these occurrences could be rare, in which case an accumulation process would still be needed.
As previously, we use $y$ to denote accumulated $\text{LLR}$ so that in the discrete case we have 
\begin{equation}
    y(t) = \sum_{s \leq t} \text{LLR}(\xi_{s}),
\end{equation}
where $\xi_s$ is the observation obtained at time $s \leq t$. Similarly, in the continuous case
\begin{equation}
    y(t) = \int_0^t \frac{dy(s)}{ds} ds
\end{equation}
where $\frac{dy}{ds}$ is given by the stochastic drift-diffusion equation described previously. 

We again assume that in a group of $N$ observers  each observer at each timestep $t$ makes an independent private observation with probability $1-c$, and all observers make a common observation with probability $c$.  Private and common observations have the same conditional distributions, $f_\pm(\xi)$ given the 
state $H^{\pm}$.

For observations drawn from such general likelihood functions, we can determine the statistics of the limiting stochastic accumulation process by averaging the impact of multiple `subobservations' on short intervals which we shrink to be infinitesimal. Focusing on a single observer $i$, define a family of stochastic processes parameterized by $k$, the number of subobservations made in an interval of length $\Delta t$. Thus, we expect the LLR increment obtained each $\Delta t$ is given
\begin{align*}
    \Delta y_t &= \sum_{l=1}^k \log \frac{f_+(\xi_{i,t}^l)}{f_-(\xi_{i,t}^l)} \\ &= {\rm E}_{\xi} \left[ \left. \sum_{l=1}^k \log \frac{f_+(\xi_{i,t}^l)}{f_-(\xi_{i,t}^l)} \right| H \right] + \left( \sum_{l=1}^k \log \frac{f_+(\xi_{i,t}^l)}{f_-(\xi_{i,t}^l)} - {\rm E}_{\xi} \left[ \left. \sum_{l=1}^k \log \frac{f_+(\xi_{i,t}^l)}{f_-(\xi_{i,t}^l)} \right| H \right] \right).
\end{align*}
We can split the sum not contained in an expectation into those observations drawn from the common pool and those not,
\begin{align*}
    \Delta y_t =& \quad {\rm E}_{\xi} \left[ \left. \sum_{l=1}^{k} \log \frac{f_+(\xi_{i,t}^l)}{f_-(\xi_{i,t}^l)} \right| H \right]\\
    &+ \left( \sum_{l=1}^{k_c} \log \frac{f_+(\xi_{i,t}^{l,c})}{f_-(\xi_{i,t}^{l,c})} + \sum_{l=1}^{k-k_c} \log \frac{f_+(\xi_{i,t}^{l,n})}{f_-(\xi_{i,t}^{l,n})} - {\rm E}_{\xi} \left[ \left. \sum_{l=1}^k \log \frac{f_+(\xi_{i,t}^l)}{f_-(\xi_{i,t}^l)} \right| H \right] \right),
\end{align*}
where $\xi_{i,t}^{l,c}$ are samples the $i^{\rm th}$ agent sees from the common pool and $\xi_{i,t}^{l,n}$ are those they see from the independent pool. For large $k$ while keeping $\Delta t$ fixed, we know the number of common observations will scale as $k_c \approx c \cdot k$, so assigning
\begin{align*}
    \pm \mu \cdot \Delta t \equiv {\rm E}_{\xi} \left[ \left. \sum_{l=1}^{k} \log \frac{f_+(\xi_{i,t}^l)}{f_-(\xi_{i,t}^l)} \right| H = H^{\pm} \right],
\end{align*}
assuming $f_{\pm}(\xi)$ are scaled appropriately as $\Delta t \to 0$. We then estimate the variability in the incremental process as $k \to \infty$ by computing
\begin{align*}
   % \left[ c \nu_{i,t}^c + (1-c) \nu_{i,t}^n \right] \cdot \Delta t &\approx 
    &\left\langle \left( \sum_{l=1}^{k_c} \log \frac{f_+(\xi_{i,t}^{l,c})}{f_-(\xi_{i,t}^{l,c})} + \sum_{l=1}^{k-k_c} \log \frac{f_+(\xi_{i,t}^{l,n})}{f_-(\xi_{i,t}^{l,n})} - {\rm E}_{\xi} \left[ \left. \sum_{l=1}^k \log \frac{f_+(\xi_{i,t}^l)}{f_-(\xi_{i,t}^l)} \right| H \right] \right)^2 \right\rangle \\
    &= \left \langle \sum_{l=1}^{k_c} \left[ \log \frac{f_+(\xi_{i,t}^{l,c})}{f_-(\xi_{i,t}^{l,c})} \right]^2 \right\rangle - c \mu^2 \cdot \Delta t^2 + \left\langle \sum_{l=1}^{k-k_c} \left[ \log \frac{f_+(\xi_{i,t}^{l,n})}{f_-(\xi_{i,t}^{l,n})} \right]^2  \right \rangle - (1-c) \mu^2 \cdot \Delta t^2 \\
    &= c \cdot {\rm Var} \left[ \sum_{l=1}^{k} \log \frac{f_+(\xi_{i,t}^{l,c})}{f_-(\xi_{i,t}^{l,c})} \right] + (1-c) \cdot {\rm Var} \left[ \sum_{l=1}^{k} \log \frac{f_+(\xi_{i,t}^{l,n})}{f_-(\xi_{i,t}^{l,n})} \right].
\end{align*}

%In the discrete \corrcase, the increment in the belief of the $i^{\text{th}}$  observer, $\Delta y_i = y_{i,t+1} - y_{i,t},$ is thus given by the LLR acting on an observation $\xi_{i,t}$
%
%\begin{equation}
%\begin{aligned}
%    \Delta y_{i,t} & =   \log{\frac{f_+(\xi_{i,t})}{f_-(\xi_{i,t})}}  %\\
%    & = (1-c) \;  \log{\frac{f_+(\xi_{i,t})}{f_-(\xi_{i,t})}}   + c \;  \log{\frac{f_+(\xi_{i,t})}{f_-(\xi_{i,t})}},
%\end{aligned}
%\end{equation}
%
%where $(1-c)$ is the probability that $\xi_{i,t}$ is obtained via an independent observation and $c$ the probability of obtaining the observation via a correlated observation.
%where we are anticipating the fraction of time this increment is drawn from an independent ($1-c$) or correlated ($c$) source. Expanding to include conditioning on the state of the environment $H$, 
%\begin{multline}
%        \Delta y_{i,t}  = (1-c)\Bigg(\mathbb{E}_\xi \Bigg[\log{\frac{f_+(\xi_{i,t})}{f_-(\xi_{i,t})}}\mid H^{\pm}\Bigg] + \log{\frac{f_+(\xi_{i,t})}{f_-(\xi_{i,t})}} - \mathbb{E}_\xi \Bigg[\log{\frac{f_+(\xi_{i,t})}{f_-(\xi_{i,t})}}\mid H^{\pm}\Bigg]\Bigg) \\
%        + c\Bigg(\mathbb{E}_\xi \Bigg[\log{\frac{f_+(\xi_{i,t})}{f_-(\xi_{i,t})}}\mid H^{\pm}\Bigg] + \log{\frac{f_+(\xi_{i,t})}{f_-(\xi_{i,t})}} - \mathbb{E}_\xi \Bigg[\log{\frac{f_+(\xi_{i,t})}{f_-(\xi_{i,t})}}\mid H^{\pm}\Bigg].
%        \label{pm}
%\end{multline}
We can thus approximate the update in the limit of rapid and infinitesimally weak observations using the Donsker Invariance Principle
\begin{equation*}
\Delta y_{i,t} \approx \pm \mu \Delta t   + \sqrt{\Delta t}(\rho_{1-c, \Delta t}(t)\eta_{1-c} + \rho_{c, \Delta t}(t)\eta_c)
\end{equation*}
where $\eta_c$ and $\eta_{1-c}$ are random variables with standard normal distributions, and 
\begin{equation}
    \begin{aligned}
        \pm \mu = \frac{1}{\Delta t}\mathbb{E}_\xi \Bigg[\log{\frac{f_+(\xi_{i,t})}{f_-(\xi_{i,t})}}\mid H^{\pm}\Bigg] ; 
        \\
        \rho_{1-c, \Delta t}^2(t)  = \frac{(1-c)}{\Delta t} \text{Var}_\xi \Bigg[\log{\frac{f_+(\xi_{i,t})}{f_-(\xi_{i,t})}}\mid H^{\pm}\Bigg]; \\
        \rho^2_{c,\Delta t}(t)  = \frac{c}{\Delta t} \text{Var}_\xi \Bigg[\log{\frac{f_+(\xi_{i,t})}{f_-(\xi_{i,t})}}\mid H^{\pm}\Bigg].
    \end{aligned}
\end{equation}
The drift $h_{\Delta t}$ and the variances $\rho^2_{c,\Delta t}$, $\rho^2_{1-c, \Delta t}$ will diverge unless $f_{\pm}(\xi)$ are properly scaled in the $\Delta t \to 0$ limit.

%We choose a specific scaling for the drift and variances arising from each observation, $\xi_{i,t}$, to ensure the limit holds. Suppose that over a time interval of duration $\Delta t$, an observation $\xi_{i,t}$ is the result of $\mu\Delta t$ separate observations. We define a family of stochastic processes parameterized by $k$, the number of subintervals into which we divide the time increment $\Delta t$. Assuming $\mu$ is large and $k>1$, each of the $k$ subintervals contains roughly $\mu_k \equiv \lfloor \mu\Delta t/k \rfloor$ observations with mean and variance that scale linearly with $\mu_k \propto \Delta t/k$. We can axieve this by approximating $\log{\frac{f_+(\xi_{t})}{f_-(\xi_{t})}}$ in Eq.~\eqref{pm} with a family of stochastic processes parameterized by $k$:
%\begin{align*}
%\Delta y_t = \sum_{l = 1}^k &\frac{\Delta t}{k}\log{\frac{f_+(\xi_l)}{f_-(\xi_l)}} + \frac{\sqrt{c\Delta t}}{\sqrt{k}}\left( \log{\frac{f_+(\xi_{l})}{f_-(\xi_{l})}} - \mathbb{E}_\xi \Bigg[\log{\frac{f_+(\xi_{l})}{f_-(\xi_{l})}}\mid H^{\pm}\Bigg]\Bigg)\right)\\
%&+ \frac{\sqrt{(1-c)\Delta t}}{\sqrt{k}}\left( \log{\frac{f_+(\xi_{l})}{f_-(\xi_{l})}} - \mathbb{E}_\xi \Bigg[\log{\frac{f_+(\xi_{l})}{f_-(\xi_{l})}}\mid H^{\pm}\Bigg]\Bigg)\right)  
%\end{align*}
%By the central limit theorem, as $k \to \infty$, the above converges in distribution to
%$$
%$$\Delta y_t \approx \Delta t h_{\Delta t}(t)  + \sqrt{\Delta t}(\rho_{1-c, \Delta t}(t)\eta_{1-c} + \rho_{c,\Delta t}(t)\eta_c)
%$$
Taking $\Delta t \to 0$ gives
\begin{equation}
    dy = \pm \mu \cdot dt + \rho_{1-c} dW_i + \rho_c  dW_c,
\end{equation}
where
\begin{align*}
    \pm \mu &= \lim_{\Delta t \to 0}h_{\Delta t}(t) = \mathbb{E}_\xi\Bigg[\log{\frac{f_+(\xi)}{f_-(\xi)}}\Big| H^{\pm}\Bigg]\\
    \rho^2_c(t) &= \lim_{\Delta t \to 0} \rho^2_{c,\Delta t}(t) = c\text{Var}_\xi\Bigg[\log{\frac{f_+(\xi)}{f_-(\xi)}}\Big|H^{\pm}\Bigg] \\
    \rho^2_{1-c}(t) &= \lim_{\Delta t \to 0} \rho^2_{c,\Delta t}(t) = (1-c)\text{Var}_\xi\Bigg[\log{\frac{f_+(\xi)}{f_-(\xi)}}\Big|H^{\pm}\Bigg].
\end{align*}
%As a concrete example, if we take 
%\begin{equation}
%    f_\pm(\xi) = \frac{1}{\sqrt{2\pi \Delta t \sigma^2}}e^{-(\xi - \Delta t \mu_\pm)/(2 \Delta t \sigma^2)}
%\end{equation}
%
%the above become (in environment $H^\pm$)
%\begin{equation}
%    \begin{aligned}
%        h(t) = \pm\frac{(\mu_+ - \mu_-)^2}{2\sigma^2} \\
%        \rho^2_{1-c} (t) = (1-c) \frac{(\mu_+ - \mu_-)^2}{\sigma^2} \\
%        \rho^2_c (t) = c \frac{(\mu_+ - \mu_-)^2}{\sigma^2}.
%    \end{aligned}
%\end{equation}

%Using $\mu_+ = 1,$ $\mu_- = -1$ and $\sigma = \sqrt{2}$, in state $H^{\pm}$ in the continuum limit the belief of agent $i$ evolves according to
%\begin{equation} \label{E:corr_continuous}
%    dy_i = dt + (\sqrt{2(1-c)}dW_i + \sqrt{2c}dW_c).
%\end{equation}
%
%This provides a correlated case extension to our previous independent case SDE, which corresponds to setting $c=0$ in Eq.~\eqref{E:corr_continuous}. 
%\begin{equation}
%    dy = dt + \sqrt{2}dW.
%\end{equation}

%In a multi-agent clique, we can specify the rate of evidence accumulation for agent $j$ with 
%\begin{equation}
%    dy_j = dt + (\sqrt{2(1-c)}dW_j + \sqrt{2c}dW_c),
%\end{equation}
%
We note that $dW_i$ corresponds to private noise, which is generated independently for each agent. The term $dW_c$ is common to all agents. %In contrast in the independent cases (or as $c \to 0^+$), the belief of each agent 
%evolved according to 
%$
%    dy_j = dt + \sqrt{2}dW_j,
%$
%and only the drift term was shared between agents.

\section{Alternative definitions of the first decider}
\label{app:1st}

In the text we defined the `first decider' as an agent chosen with equal probability from 
the set of all agents who reach threshold at the same time. Alternatively, we could pool all 
first deciders across trials, and ask for the probability that an agent in this entire pool makes a 
correct choice. In the scaling limit, the probability that multiple agents reach the threshold
at the same time converges to zero, and the two definitions are equivalent. However,
when evidence increments are finite, multiple agents can make the first decision at the same time.
In that case choosing the first decider within a trial and pooling across trials 
gives different results.

\FloatBarrier

\bibliographystyle{siam}
\bibliography{references}

\begin{thebibliography}{10}

\bibitem{banerjee1992}
{\sc A.~V. Banerjee}, {\em A simple model of herd behavior}, Q. J. Econ.,
  (1992), pp.~797--817.

\bibitem{Bogacz2006}
{\sc R.~Bogacz, E.~Brown, J.~Moehlis, P.~Holmes, and J.~D. Cohen}, {\em {The
  physics of optimal decision making: A formal analysis of models of
  performance in two-alternative forced-choice tasks.}}, Psychological Review,
  113 (2006), pp.~700--765.

\bibitem{bogacz2010}
{\sc R.~Bogacz, E.-J. Wagenmakers, B.~U. Forstmann, and S.~Nieuwenhuis}, {\em
  The neural basis of the speed--accuracy tradeoff}, Trends in neurosciences,
  33 (2010), pp.~10--16.

\bibitem{boland1989majority}
{\sc P.~J. Boland}, {\em Majority systems and the condorcet jury theorem},
  Journal of the Royal Statistical Society: Series D (The Statistician), 38
  (1989), pp.~181--189.

\bibitem{bose2017collective}
{\sc T.~Bose, A.~Reina, and J.~A. Marshall}, {\em Collective decision-making},
  Current opinion in behavioral sciences, 16 (2017), pp.~30--34.

\bibitem{Caginalp2017}
{\sc R.~J. Caginalp and B.~Doiron}, {\em Decision dynamics in groups with
  interacting members}, SIAM Journal on Applied Dynamical Systems, 16 (2017),
  pp.~1543--1562.

\bibitem{ccelen2004observational}
{\sc B.~{\c{C}}elen and S.~Kariv}, {\em Observational learning under imperfect
  information}, Games and Economic behavior, 47 (2004), pp.~72--86.

\bibitem{chittka2003bees}
{\sc L.~Chittka, A.~G. Dyer, F.~Bock, and A.~Dornhaus}, {\em Bees trade off
  foraging speed for accuracy}, Nature, 424 (2003), pp.~388--388.

\bibitem{chittka2009speed}
{\sc L.~Chittka, P.~Skorupski, and N.~E. Raine}, {\em Speed--accuracy tradeoffs
  in animal decision making}, Trends in ecology \& evolution, 24 (2009),
  pp.~400--407.

\bibitem{condorcet}
{\sc M.~d. Condorcet}, {\em Essay on the application of analysis to the
  probability of majority decisions}, Paris: Imprimerie Royale,  (1785).
\newblock Reprinted in Condorcet: Selected Writings, Keith Michael Baker, ed,
  1976.

\bibitem{conradt2005}
{\sc L.~Conradt and T.~J. Roper}, {\em Consensus decision making in animals},
  Trends in ecology \& evolution, 20 (2005), pp.~449--456.

\bibitem{Couzin2009}
{\sc I.~D. Couzin}, {\em Collective cognition in animal groups}, Trends in
  cognitive sciences, 13 (2009), pp.~36--43.

\bibitem{de2014essai}
{\sc N.~De~Condorcet}, {\em Essai sur l'application de l'analyse {\`a} la
  probabilit{\'e} des d{\'e}cisions rendues {\`a} la pluralit{\'e} des voix},
  Cambridge University Press, 2014.

\bibitem{enke19}
{\sc B.~Enke and F.~Zimmermann}, {\em Correlation neglect in belief formation},
  The Review of Economic Studies, 86 (2019), pp.~313--332.

\bibitem{galton1907vox}
{\sc F.~Galton}, {\em Vox populi},  (1907).

\bibitem{gardiner2009handbook}
{\sc C.~W. Gardiner}, {\em Stochastic methods: A handbook for the natural and
  social sciences}, vol.~4, springer Berlin, 2009.

\bibitem{gerber2009does}
{\sc A.~S. Gerber, D.~Karlan, and D.~Bergan}, {\em Does the media matter? a
  field experiment measuring the effect of newspapers on voting behavior and
  political opinions}, American Economic Journal: Applied Economics, 1 (2009),
  pp.~35--52.

\bibitem{Gold02}
{\sc J.~I. Gold and M.~N. Shadlen}, {\em Banburismus and the brain: decoding
  the relationship between sensory stimuli, decisions, and reward}, Neuron, 36
  (2002), pp.~299--308.

\bibitem{gold2007neural}
\leavevmode\vrule height 2pt depth -1.6pt width 23pt, {\em The neural basis of
  decision making}, Annu. Rev. Neurosci., 30 (2007), pp.~535--574.

\bibitem{Gold2007}
\leavevmode\vrule height 2pt depth -1.6pt width 23pt, {\em The neural basis of
  decision making}, Annu. Rev. Neurosci., 30 (2007), pp.~535--574.

\bibitem{kao2014decision}
{\sc A.~B. Kao and I.~D. Couzin}, {\em Decision accuracy in complex
  environments is often maximized by small group sizes}, Proceedings of the
  Royal Society B: Biological Sciences, 281 (2014), p.~20133305.

\bibitem{karamched2020heterogeneity}
{\sc B.~Karamched, M.~Stickler, W.~Ott, B.~Lindner, Z.~P. Kilpatrick, and
  K.~Josi{\'c}}, {\em Heterogeneity improves speed and accuracy in social
  networks}, Physical Review Letters, 125 (2020), p.~218302.

\bibitem{Karamched20}
{\sc B.~Karamched, S.~Stolarczyk, Z.~P. Kilpatrick, and K.~Josi\'{c}}, {\em
  Bayesian evidence accumulation on social networks}, SIAM Journal on Applied
  Dynamical Systems, 19 (2020), pp.~1884--1919.

\bibitem{lo2022wisdom}
{\sc A.~W. Lo and R.~Zhang}, {\em The wisdom of crowds versus the madness of
  mobs: An evolutionary model of bias, polarization, and other challenges to
  collective intelligence}, Collective Intelligence, 1 (2022),
  p.~26339137221104785.

\bibitem{marshall2017individual}
{\sc J.~A. Marshall, G.~Brown, and A.~N. Radford}, {\em Individual
  confidence-weighting and group decision-making}, Trends in ecology \&
  evolution, 32 (2017), pp.~636--645.

\bibitem{mensi2013correlations}
{\sc W.~Mensi, M.~Beljid, A.~Boubaker, and S.~Managi}, {\em Correlations and
  volatility spillovers across commodity and stock markets: Linking energies,
  food, and gold}, Economic Modelling, 32 (2013), pp.~15--22.

\bibitem{moreno2010decision}
{\sc R.~Moreno-Bote}, {\em Decision confidence and uncertainty in diffusion
  models with partially correlated neuronal integrators}, Neural computation,
  22 (2010), pp.~1786--1811.

\bibitem{mossel2014opinion}
{\sc E.~Mossel and O.~Tamuz}, {\em Opinion exchange dynamics}, arXiv preprint
  arXiv:1401.4770,  (2014).

\bibitem{newsome1989neuronal}
{\sc W.~T. Newsome, K.~H. Britten, and J.~A. Movshon}, {\em Neuronal correlates
  of a perceptual decision}, Nature, 341 (1989), pp.~52--54.

\bibitem{nitzan1982optimal}
{\sc S.~Nitzan and J.~Paroush}, {\em Optimal decision rules in uncertain
  dichotomous choice situations}, International Economic Review,  (1982),
  pp.~289--297.

\bibitem{olfati2006belief}
{\sc R.~Olfati-Saber, E.~Franco, E.~Frazzoli, and J.~S. Shamma}, {\em Belief
  consensus and distributed hypothesis testing in sensor networks}, in
  Networked Embedded Sensing and Control, Springer, 2006, pp.~169--182.

\bibitem{pratt2002quorum}
{\sc S.~C. Pratt, E.~B. Mallon, D.~J. Sumpter, and N.~R. Franks}, {\em Quorum
  sensing, recruitment, and collective decision-making during colony emigration
  by the ant leptothorax albipennis}, Behavioral Ecology and Sociobiology, 52
  (2002), pp.~117--127.

\bibitem{Ratcliff1978theory}
{\sc R.~Ratcliff}, {\em A theory of memory retrieval.}, Psychological review,
  85 (1978), p.~59.

\bibitem{Ratcliff2008}
{\sc R.~Ratcliff and G.~McKoon}, {\em The diffusion decision model: theory and
  data for two-choice decision tasks}, Neural computation, 20 (2008),
  pp.~873--922.

\bibitem{redner2001}
{\sc S.~Redner}, {\em A guide to first-passage processes}, Cambridge university
  press, 2001.

\bibitem{reina2022asynchrony}
{\sc A.~Reina, T.~Bose, V.~Srivastava, and J.~A. Marshall}, {\em Asynchrony
  rescues statistically-optimal group decisions from information cascades
  through emergent leaders}, bioRxiv,  (2022).

\bibitem{seeley1991collective}
{\sc T.~D. Seeley, S.~Camazine, and J.~Sneyd}, {\em Collective decision-making
  in honey bees: how colonies choose among nectar sources}, Behavioral Ecology
  and Sociobiology, 28 (1991), pp.~277--290.

\bibitem{shan2019family}
{\sc H.~Shan, R.~Moreno-Bote, and J.~Drugowitsch}, {\em Family of closed-form
  solutions for two-dimensional correlated diffusion processes}, Physical
  Review E, 100 (2019), p.~032132.

\bibitem{simmons2011intuitive}
{\sc J.~P. Simmons, L.~D. Nelson, J.~Galak, and S.~Frederick}, {\em Intuitive
  biases in choice versus estimation: Implications for the wisdom of crowds},
  Journal of Consumer Research, 38 (2011), pp.~1--15.

\bibitem{surowiecki2005wisdom}
{\sc J.~Surowiecki}, {\em The wisdom of crowds}, Anchor, 2005.

\bibitem{swets1961decision}
{\sc J.~A. Swets, W.~P. Tanner~Jr, and T.~G. Birdsall}, {\em Decision processes
  in perception.}, Psychological review, 68 (1961), p.~301.

\bibitem{uchida2003speed}
{\sc N.~Uchida and Z.~F. Mainen}, {\em Speed and accuracy of olfactory
  discrimination in the rat}, Nature neuroscience, 6 (2003), pp.~1224--1229.

\bibitem{Usher2001}
{\sc M.~Usher and J.~L. McClelland}, {\em The time course of perceptual choice:
  the leaky, competing accumulator model.}, Psychological review, 108 (2001),
  p.~550.

\bibitem{valone1989group}
{\sc T.~J. Valone}, {\em Group foraging, public information, and patch
  estimation}, Oikos,  (1989), pp.~357--363.

\bibitem{veliz16}
{\sc A.~Veliz-Cuba, Z.~P. Kilpatrick, and K.~Josi{\'c}}, {\em Stochastic models
  of evidence accumulation in changing environments}, SIAM Review, 58 (2016),
  pp.~264--289.

\bibitem{wald1945}
{\sc A.~Wald}, {\em Sequential tests of statistical hypotheses}, The annals of
  mathematical statistics, 16 (1945), pp.~117--186.

\bibitem{Wald1948}
{\sc A.~Wald and J.~Wolfowitz}, {\em Optimum character of the sequential
  probability ratio test}, The Annals of Mathematical Statistics,  (1948),
  pp.~326--339.

\end{thebibliography}
%*****************************************************************
\end{document}